\documentclass[sigconf,10pt]{acmart} 

\setcopyright{none}
\renewcommand\footnotetextcopyrightpermission[1]{}
\usepackage{pifont}
\usepackage{subcaption}
\usepackage{multirow}
\usepackage{booktabs}
\usepackage{enumitem} 
\usepackage[most]{tcolorbox}

\usepackage{microtype}
\usepackage{enumitem}
\setlist[itemize]{topsep=0.5pt,itemsep=0pt,parsep=0pt,partopsep=0pt,leftmargin=*}
\setlist[enumerate]{topsep=0.5pt,itemsep=0pt,parsep=0pt,partopsep=0pt,leftmargin=*}

\newcommand{\cmark}{\ding{51}} 
\newcommand{\xmark}{\ding{55}} 
\newcommand{\pmark}{\ensuremath{\triangle}} 

\begin{document}

\title{QoEReasoner: An Agentic Reasoning Framework for Automated and Explainable QoE Diagnosis in RANs}

\author{Qizhe Li}
\authornote{Both authors contributed equally to this research.}
\email{qizheli@link.cuhk.edu.cn}
\affiliation{%
  \institution{The Chinese University of Hong Kong, Shenzhen}
  \city{Shenzhen}
  \country{China}
}
\affiliation{%
  \institution{Shenzhen Research Institute of Big Data}
  \city{Shenzhen}
  \country{China}
}

\author{Haolong Chen}
\authornotemark[1]
\email{haolongchen1@link.cuhk.edu.cn}
\affiliation{%
  \institution{The Chinese University of Hong Kong, Shenzhen}
  \city{Shenzhen}
  \country{China}
}
\affiliation{%
  \institution{Shenzhen Research Institute of Big Data}
  \city{Shenzhen}
  \country{China}
}

\author{Shan Dai}
\email{daishan@cuhk.edu.cn}
\affiliation{%
  \institution{Shenzhen Research Institute of Big Data}
  \city{Shenzhen}
  \country{China}
}

\author{Zhuo Li}
\email{zhuoli3@link.cuhk.edu.cn}
\affiliation{%
  \institution{The Chinese University of Hong Kong, Shenzhen}
  \city{Shenzhen}
  \country{China}
}
\affiliation{%
  \institution{Shenzhen Research Institute of Big Data}
  \city{Shenzhen}
  \country{China}
}

\author{Zhiwei Hu}
\email{zhiweihu@link.cuhk.edu.cn}
\affiliation{%
  \institution{The Chinese University of Hong Kong, Shenzhen}
  \city{Shenzhen}
  \country{China}
}
\affiliation{%
  \institution{Shenzhen Research Institute of Big Data}
  \city{Shenzhen}
  \country{China}
}

\author{Xuan Li}
\email{lixuan69@huawei.com}
\affiliation{%
  \institution{Huawei Technologies Co., Ltd.}
  \city{Shenzhen}
  \country{China}
}

\author{Guangxu Zhu}
\authornote{Corresponding author.}
\email{gxzhu@sribd.cn}
\affiliation{%
  \institution{Shenzhen Research Institute of Big Data}
  \city{Shenzhen}
  \country{China}
}

\author{Qingjiang Shi}
\email{shiqj@tongji.edu.cn}
\affiliation{%
  \institution{Shenzhen Research Institute of Big Data}
  \city{Shenzhen}
  \country{China}
}
\affiliation{%
  \institution{Tongji University}
  \city{Shanghai}
  \country{China}
}

\renewcommand{\shortauthors}{Li et al.}

\begin{abstract}
Diagnosing Quality-of-Experience (QoE) degradations in operational Radio Access Networks (RANs) is a critical but notoriously complex task, traditionally requiring labor-intensive expert analysis over high-dimensional, cross-layer telemetry. While Large Language Models (LLMs) offer unprecedented reasoning capabilities, they are fundamentally unsuited for raw RANs troubleshooting: they fail at numeric time-series analysis, hallucinate protocol-violating causal links, and lack the stateful rigor required for multi-step fault localization. To bridge this gap, we present QoEReasoner, an end-to-end, LLM-driven agentic system designed for automated and explainable QoE diagnosis. QoEReasoner tames the inherent unpredictability of LLMs by grounding their reasoning in the physical realities of the network. It employs deterministic tools to reliably translate raw numeric KPIs into structured evidence, enforces protocol-consistent fault propagation through a domain-specific Knowledge Base, and leverages a Historical Bank of expert-validated cases to guide hypothesis generation. A stateful central planner orchestrates this closed-loop process across anomaly detection, causal tracing, and root-cause localization. Evaluations on real-world operational RANs datasets demonstrate that QoEReasoner outperforms strong baselines by 18\%–40\% in  accuracy across multiple diagnostic tasks. Furthermore, it reduces diagnostic time from approximately 30 minutes of manual expert analysis to just 3 minutes per session, delivering highly interpretable, expert-grade reports while remaining robust across diverse LLM backbones.

\end{abstract}



\maketitle


\begin{figure}[t]
    \centering
    \includegraphics[width=1\linewidth]{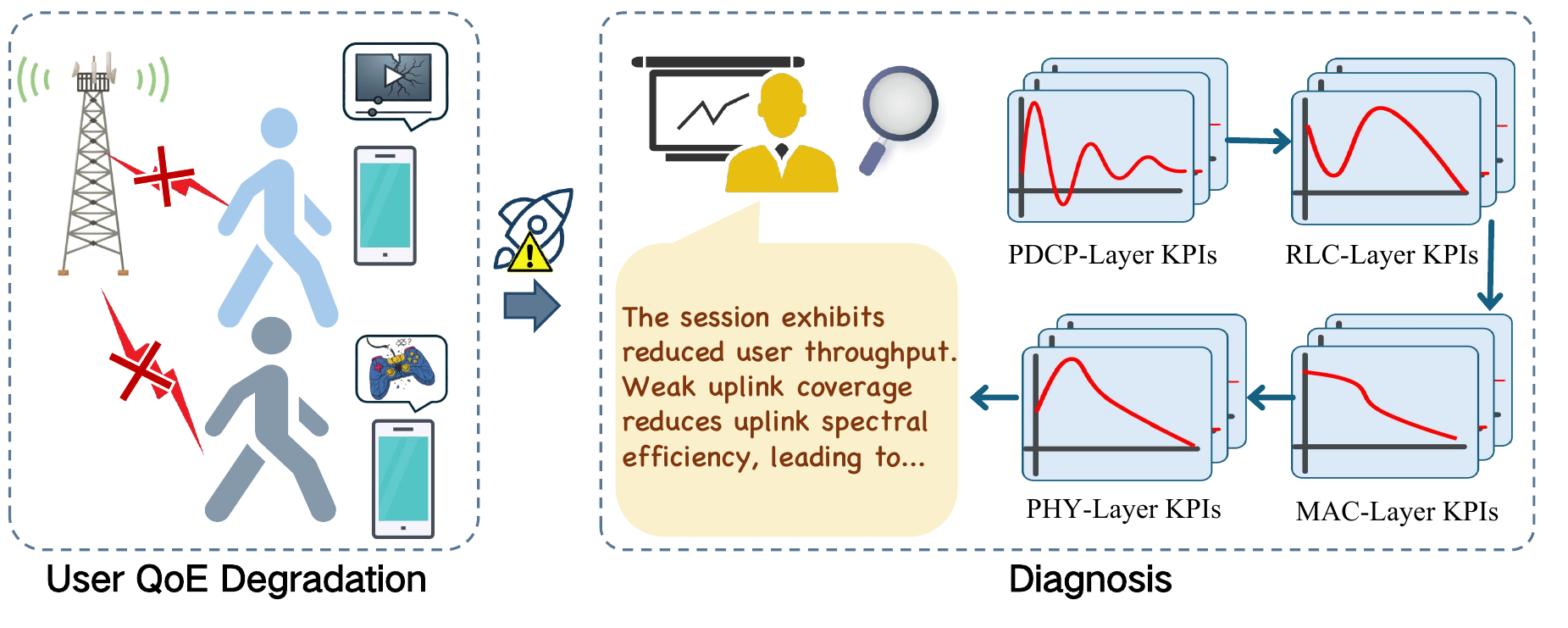}
    \caption{Illustration of QoE diagnosis in operational RANs.
    }
    \label{fig:task_illus}
\end{figure}

\section{Introduction}
\label{sec:intro}



Maintaining a high user Quality-of-Experience (QoE) is the fundamental objective of modern Radio Access Networks (RANs)~\cite{alam2025comprehensive, herrera2025tutorial, boateng2025survey}. With the rapid scaling of mobile networks, the explosive growth in user sessions has made it impossible to manually track network health, elevating the automated detection of QoE degradations to a critical priority. Yet, simply flagging a degraded session is insufficient for network optimization. To actually fix the network, operators must diagnose why the degradation occurred. As shown in Fig.~\ref{fig:task_illus}, this requires working backward from user-perceived symptoms through a maze of high dimentional, time-varying, cross-layer Key Performance Indicators (KPIs) to infer root causes such as interference, weak coverage, or cell overload~\cite{khoramnejad2025generative, osee2025quality}. Since this troubleshooting process is so labor-intensive that practical QoE management in operational RANs demands a framework that is both \textit{automated} in its execution and \textit{explainable} in its reasoning.

In order to realize such a holistic QoE diagnosis pipeline, an effective candidate system should satisfy the following three key non-negotiable requirements. First, to interpret massive and heterogeneous KPI telemetry, the system needs \textbf{(R1) reliable numeric perception} to transform raw measurements into trustworthy representations of network state, in order to achieve effective detection of QoE-degraded sessions. Second, beyond mere detection, the system should provide \textbf{(R2) causally grounded cross-layer explanations}, which trace how impairments propagate across protocol layers and thus connect observed symptoms to root causes through protocol-consistent causal paths.
Third, since QoE diagnosis inherently spans fragmented yet interdependent tasks from anomaly detection and causal tracing to root-cause localization, the system demands \textbf{(R3) coherent cross-task reasoning} to enforce logical consistency across the entire diagnostic context. The three requirements capture the essential properties of a practical QoE diagnosis system.



However, existing approaches still fail to jointly satisfy R1--R3.
Prior QoE diagnosis in RANs has been primarily studied through heuristic and deep learning (DL) approaches. Specifically, the former relies on expert-defined rules and transparent threshold-based diagnosis workflows~\cite{ciocarlie2014feasibility, moulay2020novel, yuan2020anomaly}, which provide basic interpretability for degraded-session identification, partially addressing R1. However, rule-based workflows remain brittle to context variation and incomplete observations, due to the heavy dependent on manual rule engineering and domain expertise~\citep{yuan2020anomaly}. Such limitations make heuristic methods insufficient for deeper causal diagnosis tasks, especially causally grounded cross-layer explanations (R2) and coherent cross-task reasoning (R3). 
Although \textit{DL-based approaches} successfully address QoE anomaly detection and root-cause classification by mapping network observations with user experience (R1) in a data-driven way~\cite{chawla2020interpretable, sun2024spotlight, li2025dk, ghuge2023deep}, most of DL predictions remain Black-Box, without the protocol-aware transparency required for causal explanations (R2). Moreover, most DL solutions treat each task in QoE diagnosis as independent task, which cannot maintain the logical continuity necessary to resolve the complex, multi-stage reasoning demanded by R3.

Large Language Models (LLMs) have recently shown promise for wireless-network intelligence~\cite{liu2025deepseek, yang2025qwen3, shen2025autoiot, wang2025large, chen2025first}, including network-management tasks such as network slicing~\cite{tong2025wirelessagent, tong2026wirelessagent}. 
However, their direct application to QoE diagnosis remains challenging despite the ability to synthesize scattered evidence, articulate causal hypotheses, and generate human-interpretable reasoning through natural-language interaction. Specifically, LLMs struggle with numeric-intensive, high-dimensional RANs data due to unreliable numerical understanding and prompt sensitivity~\cite{zhou2025can, park2025revisiting}, failing to reliably satisfy R1.\footnote{We discuss this limitation in \ref{subsec:preliminary}.}
Furthermore, without expert constraints, inherent hallucinations may contradict domain knowledge\cite{lee2025verisafe, chen2026overview}, leading to violation of R2 and R3. Consequently, as Table~\ref{tab:capability_comparison} summarizes, existing paradigms fail to jointly satisfy R1--R3.

\begin{table}
  \centering
  \caption{Capability comparison of QoE diagnosis paradigms
  (\cmark: supported; \pmark: partial; \xmark: not supported).}
  \label{tab:capability_comparison}
  \resizebox{\columnwidth}{!}{%
  \setlength{\tabcolsep}{3pt}
  \fontsize{8.5pt}{11pt}\selectfont
  \begin{tabular}{l c c c c}
    \toprule
    \textbf{Requirement}
    & \textbf{Heuristic}
    & \textbf{DL}
    & \textbf{LLM}
    & \textbf{\underline{\textit{QoEReasoner}}} \\
    \midrule
    (R1) Reliable numeric perception
      & \pmark
      & \cmark
      & \xmark
      & \cmark \\
    (R2) Causally grounded explanation
      & \xmark
      & \xmark
      & \pmark
      & \cmark \\
    (R3) Coherent cross-task reasoning
      & \xmark
      & \xmark
      & \pmark
      & \cmark \\
    \bottomrule
  \end{tabular}
  }
\end{table}

Motivated by this gap, we present \textit{\textbf{QoEReasoner}}, the first LLM-based agentic diagnostic framework designed for automated and explainable QoE analysis that addresses R1---R3 simultaneously.
By orchestrating external tools invocation with closed-loop observation, decisions making and actions taking, \textit{{QoEReasoner}} seamlessly integrates the golden expert priors from heuristic methods, stable predictive power from deep learning, and interpretable reasoning abilities from LLMs. 


Specifically, for an input KPI session, an LLM-based \emph{Planner} invokes \emph{KPI Perception} to characterize the QoE state through trend analysis and degradation recognition (R1). Upon confirming reliable degradation, the \emph{Planner} advances to \emph{Fault Causal Chain Reasoning} to infer plausible root causes via tracing protocol-consistent fault propagation paths (R2). The verified diagnostic findings are then forwarded to the \emph{Reporter} for diagnosis report generation (R3). To support this diagnostic process, \emph{QoEReasoner} also incorporates several auxiliary components: (1) a \emph{Tool Pool} that provides functional interfaces for tasks (e.g., KPI preprocess and inference with pre-trained anomaly classification models); (2) a \emph{Knowledge Base} that contains domain knowledge (e.g., heuristic rules and causal constraints), offering guidance and constraints; (3) a \emph{Historical Bank} supplies expert-validated reference cases to inform prior-guided fault diagnosis. Comprehensive experiments on real-world mobile network datasets show that \textit{QoEReasoner} outperforms strong baselines by 18\%--40\% across multiple diagnostic tasks, significantly reduces diagnosis time from about 30 minutes of manual expert analysis to around 3 minutes per session, stably produces protocol-consistent and evidence-grounded explanations, and consistently remains robust across diverse LLM backbones from different model families and parameter scales. We summarize our contributions as follows:

\begin{itemize}[itemsep=0em, topsep=0em, leftmargin=1em]

\item To the best of our knowledge, \textit{QoEReasoner} is the first agentic framework for automated multi-task QoE diagnosis in mobile networks, unifying anomaly detection, causal tracing, and root-cause localization.

\item We design and implement several novel modules, including \emph{KPI Perception}, \emph{Fault Causal Chain Reasoning}, and the \emph{Historical Bank}, to enable explainable diagnosis through evidence-grounded causal reasoning.\footnote{To support reproducibility, we will release the implementation code and accompanying documentation upon publication.}

\item We conduct comprehensive evaluations on real-world mobile network datasets, showing that \textit{QoEReasoner} consistently outperforms prior approaches across multiple diagnostic tasks and remains robust across diverse LLM backbones.

\end{itemize}

\section{Problem Formulation and Preliminary Analysis}
We first formalize the QoE diagnosis task researched in this paper, and provide a preliminary discussion about why naive LLM reasoning struggles in achieving reliable RANs diagnoses, which directly motivates the design principles behind our proposed \textit{QoEReasoner}.

\subsection{Task Formulation}
\label{sec:task formu}

\begin{figure}
  \centering
  \includegraphics[width=\linewidth]{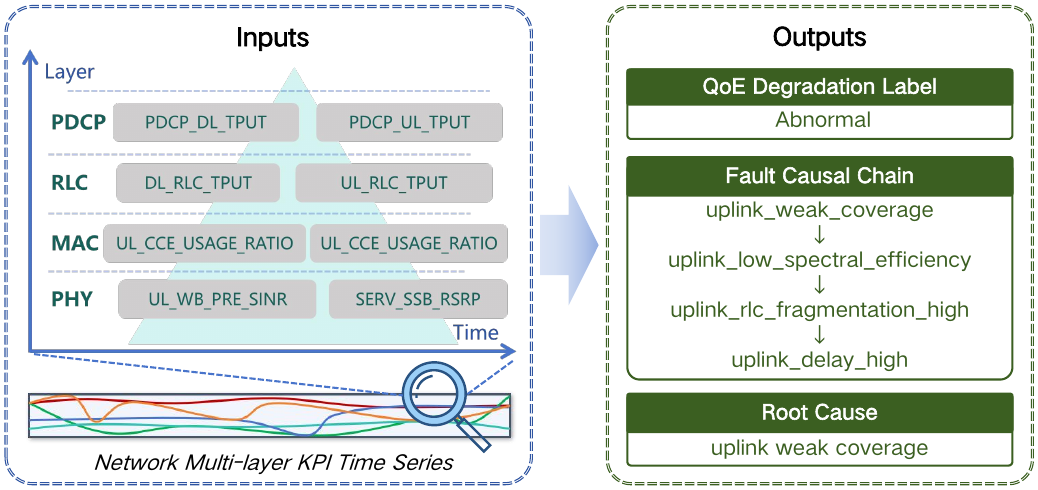}
  \caption{
  Illustration of the QoE diagnosis tasks.
  }
  \label{fig:label_example}
\end{figure}


The input of the QoE diagnostic tasks is a session-level multivariate time-series data collected from the physical (PHY), medium access control (MAC), radio link control (RLC), and packet data convergence protocol (PDCP) layers, called KPI telemetry, formulated as
\begin{equation}
  \mathbf{X} \in \mathbb{R}^{T \times M},
\end{equation}
where $T$ denotes the number of timestamps and $M$ denotes the number of KPI dimensions.
We define the diagnostic tasks (illustrated in Fig.~\ref{fig:label_example}) as follows:

\paragraph{Anomaly Detection (AD)} that aims to determine whether a session exhibits user-perceived QoE degradation.
Let $y^{\mathrm{deg}} \in \{0,1\}$ denote the degradation label, where $y^{\mathrm{deg}}=1$ indicates that the session is annotated as degraded.
Given session measurements $\mathbf{X}$, the task is formulated as a binary classification problem:
\begin{equation}
  f_{\text{AD}}: \mathbf{X} \mapsto \hat{y}^{\mathrm{deg}}, \quad \hat{y}^{\mathrm{deg}} \in \{0,1\}.
\end{equation}

\paragraph{Fault Chain Reasoning (FCR) } that aims to infer a structured, directed causal chain that captures the propagation of anomalies across protocol layers toward the observed QoE degradation, as illustrated in Fig.~\ref{fig:label_example}.
\begin{equation}
  \mathbf{c}: c_1 \rightarrow c_2 \rightarrow \cdots \rightarrow c_L,
\end{equation}
where each node $c_i \in \mathcal{C}$ denotes an anomaly phenomenon from a global vocabulary $\mathcal{C}$, and each edge $c_i \rightarrow c_{i+1}$ denotes a protocol-consistent causal dependency. Given $\mathbf{X}$, we formulate FCR task as:
\begin{equation}
  f_{\text{FCR}}: \mathbf{X} \mapsto \hat{\mathbf{c}} = (\hat{c}_1,\ldots,\hat{c}_{\hat{L}}), \quad \hat{c}_i \in \mathcal{C},
\end{equation}
where $\hat{\mathbf{c}}$ is the inferred causal path linking lower-layer abnormalities to the final QoE degradation.

\paragraph{Root-Cause Categorization (RCC)} aims to identify the category of the root cause responsible for the observed QoE degradation. 
Let $y^{\mathrm{rc}} \in \{1,\ldots,K\}$ denote the root-cause category label, where each class corresponds to a predefined category of degradation factors, such as interference and weak coverage. Given session measurements $\mathbf{X}$, the task is formulated as a $K$-class classification problem:
\begin{equation}
  f_{\text{RCC}}: \mathbf{X} \mapsto \hat{y}^{\mathrm{rc}}, \quad \hat{y}^{\mathrm{rc}} \in \{1,\ldots,K\}.
\end{equation}



\subsection{Preliminary Experiments}
\label{subsec:preliminary}
\begin{figure}
  \centering
  \includegraphics[width=\linewidth]{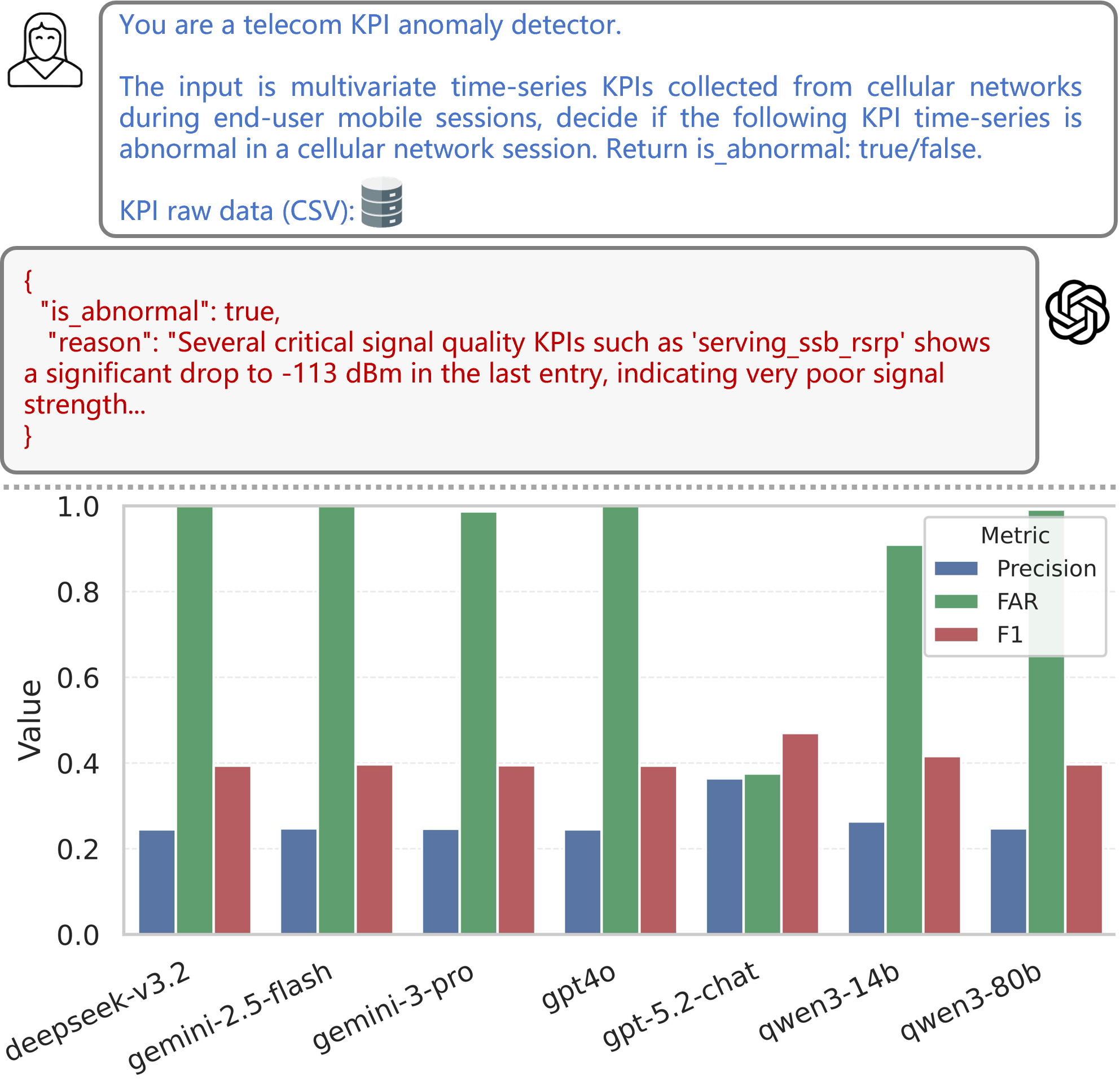}
  \caption{Prompt-based anomaly detection with LLMs: (top) example prompt for KPI anomaly detection; (bottom) performance comparison across LLMs measured by precision, false alarm rate (FAR), and F1 score.}
  \label{fig:llm_weak}
\end{figure}

Although LLMs are effective for reasoning~\cite{kong2025timemqa,liu2025large,wu2024netllm, maatouk2024tele}, it remains unclear whether they can reliably interpret RANs KPIs time-series while producing stable judgments.
To probe this capability gap, we conduct a preliminary study on prompt-based KPI anomaly detection.\footnote{Detailed settings are provided in Appendix~\ref{appen:prelim_llm_setup}.}

As illustrated in Fig.~\ref{fig:llm_weak}, prompt-based anomaly detection exhibits substantial performance variance across models. More importantly, none of the evaluated LLMs provides both accurate and stable predictions, indicating inherent limitations in directly applying general-purpose LLMs to numeric RANs diagnosis tasks.

\paragraph{Limitations in Understanding Numeric KPI Telemetry.}
QoE diagnosis in RANs relies on high-dimensional multivariate time series spanning multiple protocol layers, involving over ninety correlated KPIs~\cite{yuan2020anomaly, sun2024spotlight, li2025dk}. Such inputs require fine-grained temporal pattern recognition and cross-layer semantic grounding, yet general-purpose LLMs are primarily optimized for textual and symbolic reasoning and remain weak at structured time-series understanding~\cite{zhou2025can}. Moreover, many degradation scenarios in operational RANs are unlikely to be well covered in LLM pretraining corpora, which further limits generalization in this task. Additionally, the cues for anomaly detection and root-cause identification are often subtle and distributed across both temporal and feature dimensions, further making prompt-only inference unreliable.

\paragraph{Conservative Bias and Decision Instability.}
Empirically, we observe that LLMs often achieve high recall at the cost of excessive false alarms, aligning with the \emph{conservative bias} phenomenon~\cite{sriramanan2024llm, aguda2025conservative}. In our AD task setting, this bias manifests as a tendency to label cases as anomalous once the detection task is posed, even when the evidence is ambiguous.
This reflects that LLMs lack explicit distributional grounding and calibrated uncertainty estimation~\cite{reif2024beyond}, producing semantically plausible predictions rather than conservative null decisions.
We also observe pronounced prompt sensitivity where minor wording variations lead to substantially different predictions, further undermining reliability.


We argue that the limitations of directly applying pretrained LLMs will extend to other diagnosis-related tasks in RANs, thereby motivating us to adopt an LLM-based agentic paradigm for automatic and explainable QoE diagnosis, in which the LLM is augmented with external modules.

\section{Overview of Key Ideas}
\label{subsec:key-ideas}

In this paper, we develop an agentic system called \textit{QoEReasoner} for QoE diagnosis in RANs. Built upon several key design ideas, the system integrates multiple novel modules with the reasoning and orchestration capabilities of the LLM, enabling the agent to satisfy R1--R3.

\paragraph{Overcoming Numeric Weakness via External Tools $\rightarrow$ (R1): }
We perform network KPI awareness by introducing external tools, such as data preprocessing modules, that convert raw numerical metrics into text-friendly evidence for the LLM. This tool-grounded design alleviates the limitations of LLMs in numerical understanding, as shown in \S\ref{subsec:preliminary}. Compared with fine-tuning the LLM for task-specific adaptation, we leverage external tools as a more practical alternative to produce reliable and reproducible outputs.

\paragraph{Preventing Hallucinations via Knowledge Constraints $\rightarrow$ (R2): }
Diagnostic reasoning is restricted to a constrained hypothesis space defined by protocol semantics and fault causal chains are retrieved and verified link-by-link against \emph{Knowledge Base (KB)} constraints. The KB organizes valid cross-layer propagation patterns as a global causal graph, thereby substantially narrowing the effective hypothesis space and limiting unsupported or protocol-inconsistent causal reasoning during diagnosis.

\paragraph{Guiding Diagnosis via Verified Historical Cases $\rightarrow$ (R2): }
Cases in \emph{Historical Bank (HB)} serve as soft priors for hypothesis prioritization, but all retrieved hypotheses must pass evidence-based re-validation before acceptance.
The HB transforms fault causal chain construction from open-ended combinatorial search into constrained hypothesis verification.

\paragraph{Ensuring Logical Consistency via a Stateful Planner $\rightarrow$ (R3):}
The planner maintains an explicit diagnostic state and iteratively decides what to inspect next, terminating only when the accumulated evidence is sufficiently consistent. Such design enables evidence-consistent coordination across multiple diagnosis-related tasks, so that intermediate information between modules can be mutually informed and progressively refined. On top of this process, the reporter generates a concise summary report by consolidating the verified diagnosis evidence.

Together, these designs fundamentally distinguish our framework from prior approaches and reshape QoE diagnosis as a structured, evidence-driven decision process.

\begin{figure*}
  \centering
  \includegraphics[width=\linewidth]{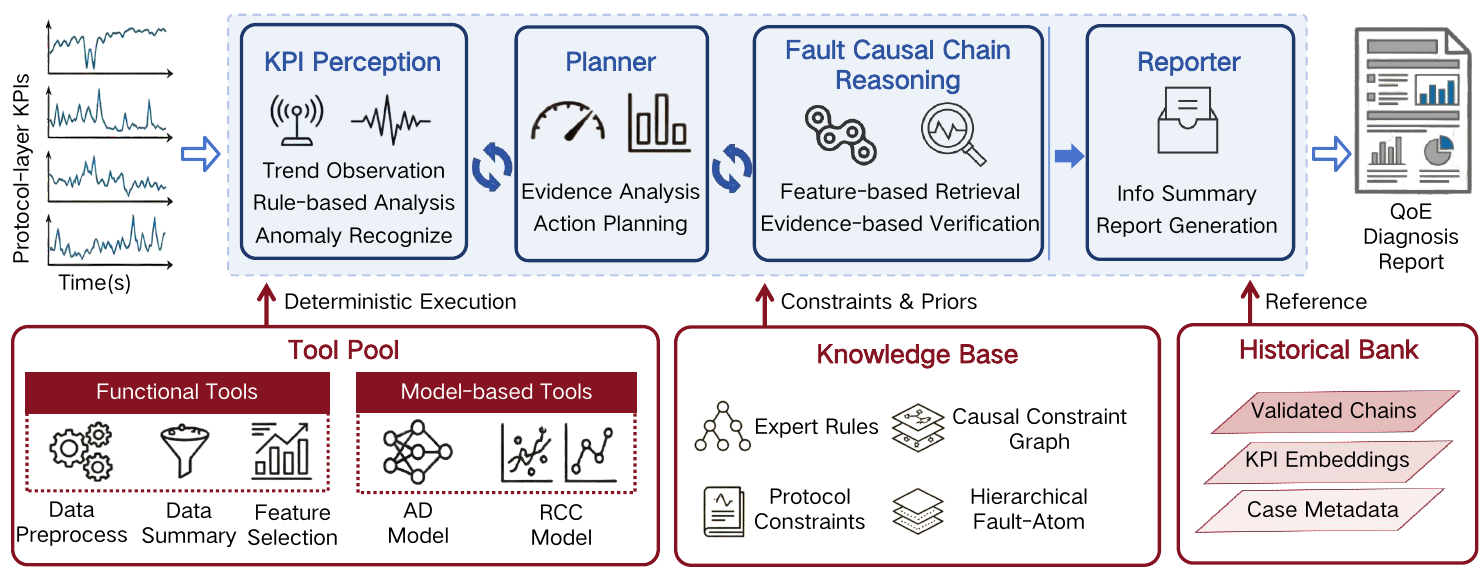}
  \caption{System overview of \textit{QoEReasoner}.}
  \label{fig:agent_architecture}
\end{figure*}



\section{System Design}
\label{sec:system-design}

As shown in Fig.~\ref{fig:agent_architecture}, \textit{QoEReasoner} adopts a modular architecture organized into two complementary groups.
The \textbf{core diagnostic modules} implement the main reasoning loop:
the \emph{planner} (\S\ref{subsec:planner}) drives the stateful closed loop by maintaining a shared diagnostic state; \emph{KPI Perception} (\S\ref{subsec:kpi-perception}) perceives network KPI status and converts raw telemetry into structured evidence; \emph{Fault Causal Chain Reasoning (FCR)} (\S\ref{subsec:rc-reasoning}) performs evidence-grounded fault propagation analysis on abnormal sessions; and the \emph{reporter} (\S\ref{subsec:reporter}) produces human-readable diagnostic reports.
The \textbf{supporting components} provide shared resources consumed by the core modules: the \emph{Tool Pool} (\S\ref{subsec:tp}) supplies deterministic analytical capabilities, the \emph{Knowledge Base (KB)} (\S\ref{subsec:kb}) enforces protocol-consistent structural constraints and provides domain knowledge required for diagnosis, and the \emph{Historical Bank (HB)} (\S\ref{subsec:historical-bank}) offers case-based references for hypothesis initialization.


\subsection{Core Diagnostic Modules}
\label{subsec:core-modules}

\subsubsection{Planner: State-driven Multi-Task Coordination}
\label{subsec:planner}
\begin{figure}
    \centering
    \includegraphics[width=\linewidth]{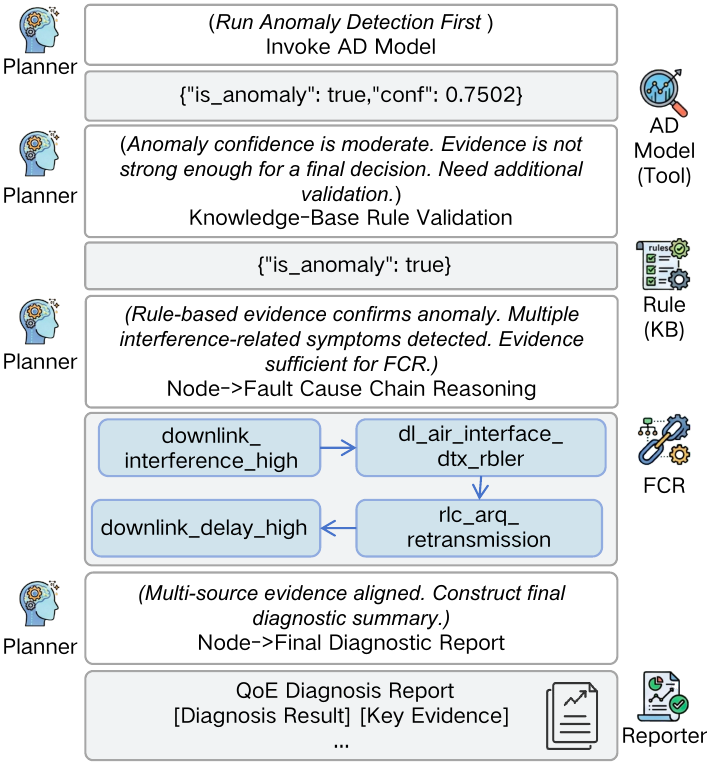}
    \caption{Example of state-driven planner for QoE diagnosis.}
    \label{fig:planner_example}
\end{figure}


The \emph{planner} serves as the control core of \textit{QoEReasoner}, as illustrated in Fig.~\ref{fig:planner_example}. Built on top of the LLM, it orchestrates the multi-step diagnosis process by selecting the next diagnostic action at each iteration. Specifically, it maintains a shared workflow state that records the message history and intermediate outputs from other modules, such as AD results, rule-based judgments, and FCR outputs. Based on this accumulated context, the planner chooses the next step from a constrained action space, such as invoking additional diagnostic nodes to gather further evidence or terminating and forwarding the collected results to the reporter. After each action, the returned information is written back into the shared state, allowing subsequent decisions to condition on the updated evidence. In this way, the planner acts as an LLM-driven controller over an explicit workflow state, enabling adaptive multi-step reasoning while keeping the overall diagnostic procedure interpretable and controllable.

\subsubsection{KPI Perception: Network State Awareness}
\label{subsec:kpi-perception}

This module provides the planner (\S\ref{subsec:planner}) with structured observations of network KPI conditions for diagnosis. Given a KPI segment, it gathers evidence under the planner’s coordination by invoking deterministic tools, consulting the \emph{Knowledge Base} (\S\ref{subsec:kb}), or using auxiliary classifiers when needed. In parallel, it summarizes protocol-layer time series using the tools in \S\ref{subsec:tp}, extracting compact descriptors such as distributional statistics, directional trends, and fluctuation indicators over normalized KPI windows. The resulting typed state objects encode the observed KPI conditions in a structured form, and are propagated through the \textit{QoEReasoner} pipeline to support downstream reasoning and report generation (\S\ref{subsec:reporter}).

\begin{figure}
    \centering

    \begin{subfigure}[t]{0.5\textwidth}
        \centering
        \includegraphics[width=\linewidth]{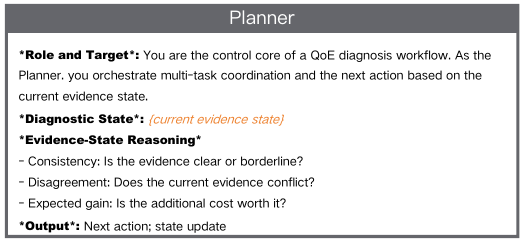}
        \caption{Planner}
        \label{fig:planner_prompt}
    \end{subfigure}
    
    \begin{subfigure}[t]{0.5\textwidth}
        \centering
        \includegraphics[width=\linewidth]{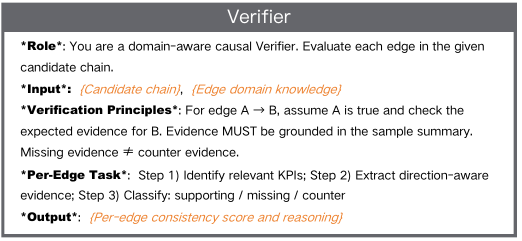}
        \caption{Verifier}
        \label{fig:verifier_prompt}
    \end{subfigure}

    \begin{subfigure}[t]{0.5\textwidth}
        \centering
        \includegraphics[width=\linewidth]{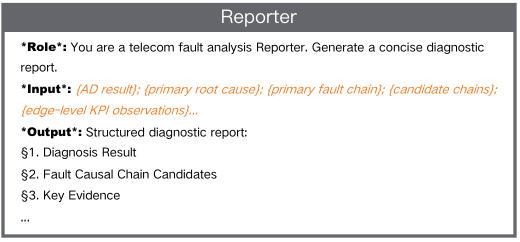}
        \caption{Reporter}
        \label{fig:reporter_prompt}
    \end{subfigure}

    \caption{The prompt template for three core LLM-driven modules.}
    \label{fig:prompt_schemas}
\end{figure}

\subsubsection{Fault Causal Chain Reasoning: Explainable QoE Degradations}
\label{subsec:rc-reasoning}

The Fault Causal Chain Reasoning (FCR) module is designed to explain, in an evidence-grounded manner, how an underlying root cause propagates across protocol layers and eventually gives rise to the observed QoE degradation. It casts fault-chain construction as a structured reasoning process that combines feature-based retrieval of historical cases with evidence-based verification of downstream symptoms along the propagation path. The final output is an ordered cause-to-effect sequence, where the first fault atom denotes the inferred root cause and each subsequent atom reflects a downstream manifestation of fault propagation.

\paragraph{Retrieval:}
Direct causal reasoning over high-dimensional KPI time series is both statistically fragile and difficult to control, especially when relying on language models alone.
To address this challenge, FCR adopts a case-based retrieval strategy that searches for historical fault instances exhibiting similar symptom patterns.
Each historical case is indexed by a vector embedding produced by the shared representation encoder in the \emph{Tool Pool} (\S\ref{subsec:tp}), ensuring representation consistency between retrieval and downstream reasoning.
Retrieval is guided by two complementary signals.
First, a semantic prior derived from the root-cause category predicted by a black-box pretrained model narrows the search space toward diagnostically relevant regions.
Second, embedding-space similarity allows structurally similar cases from other categories to be retained, preserving diversity in candidate chains.
By integrating these two sources of retrieval signals, the strategy extracts a top-$n$ candidate subset from the HB for subsequent evidence-based verification.

\paragraph{Verifier:}
Retrieved fault causal chains are aggregated into a candidate set and subjected to explicit causal verification.
Each chain is decomposed into adjacent atom pairs, corresponding to directed fault-propagation edges.
For each edge, the verifier queries the KB~(\S\ref{subsec:kb}) for expected KPI-level manifestations.
The verifier then checks the consistency between observed KPI behaviors and the expected constraints implied by each causal edge as shown in Fig.~\ref{fig:verifier_prompt}.
This process yields per-edge verification decisions as well as an overall chain-level consistency score, enabling fine-grained, auditable causal validation.

\paragraph{Adjuster:}
The final chain ranking is obtained by fusing retrieval similarity scores with causal verification scores, where the fusion weight is adaptively adjusted according to the discriminative strength of each signal, thereby mitigating over-reliance on either surface-level similarity or strict rule satisfaction alone. The top-ranked alternatives are then retained in the agent state to support subsequent planning and downstream diagnostic procedures. Within \textit{QoEReasoner}, the starting atom of the inferred causal chain is taken as the identified root cause, and the corresponding root-cause category is deterministically mapped from that atom.

By design, FCR serves as a disciplined reasoning layer that bridges data-driven signals, protocol knowledge, and interpretable diagnosis, thereby making causal explanations more reliable for network optimization.

\subsubsection{Reporter: Evidence-Grounded Report Generation}
\label{subsec:reporter}
The reporter serves as the interpretability interface for \textit{QoEReasoner}, which uses an LLM to aggregate the diagnostic state maintained by the planner, along with the supporting evidence accumulated during reasoning, and converts them into concise, human-readable reports.

\subsection{Supporting Components}
\label{subsec:supporting-components}
The core diagnostic modules rely on three shared supporting components described below.
\subsubsection{Tool Pool: Deterministic Skill Support}
\label{subsec:tp}
The Tool Pool serves as a shared skill repository for multiple modules in the framework, providing access to a heterogeneous set of capabilities through well-defined interfaces, which can be broadly grouped into two categories.

\paragraph{Functional Tools:}
These tools perform data-level operations that transform raw KPI streams into structured representations. For example, the \textit{data preprocessing tool} normalizes KPI values and interpolates missing entries to ensure numerical consistency; the \textit{data summary tool} derives statistical profiles such as percentiles and peak/valley statistics; and the \textit{feature selection tool} extracts task-relevant KPI subsets.

\paragraph{Model-Based Tools:}
These tools consist of task-specific pre-trained models with frozen parameters, providing data-driven capabilities that complement the agent's reasoning process.
For instance, the AD model is mainly used by the KPI Perception module to detect abnormal sessions and output confidence scores.
The RCC model provides data-driven priors for fault causal chain reasoning by identifying likely root-cause categories.
Meanwhile, its encoder can be reused as an embedding function to project KPI patterns into a latent root-cause space, which supports retrieval (\S\ref{subsec:rc-reasoning}) and facilitates more structurally informed hypothesis generation.

Within \textit{QoEReasoner}, tool invocation is dynamically orchestrated based on the diagnostic state and available evidence. This design decouples specialized analytical capabilities from the language-model core, allowing the framework to flexibly reuse deterministic tools while improving the reliability and controllability of diagnosis.

\subsubsection{Knowledge Base: Structured Domain Constraints}
\label{subsec:kb}

\begin{figure}
    \centering
    \includegraphics[width=1\linewidth]{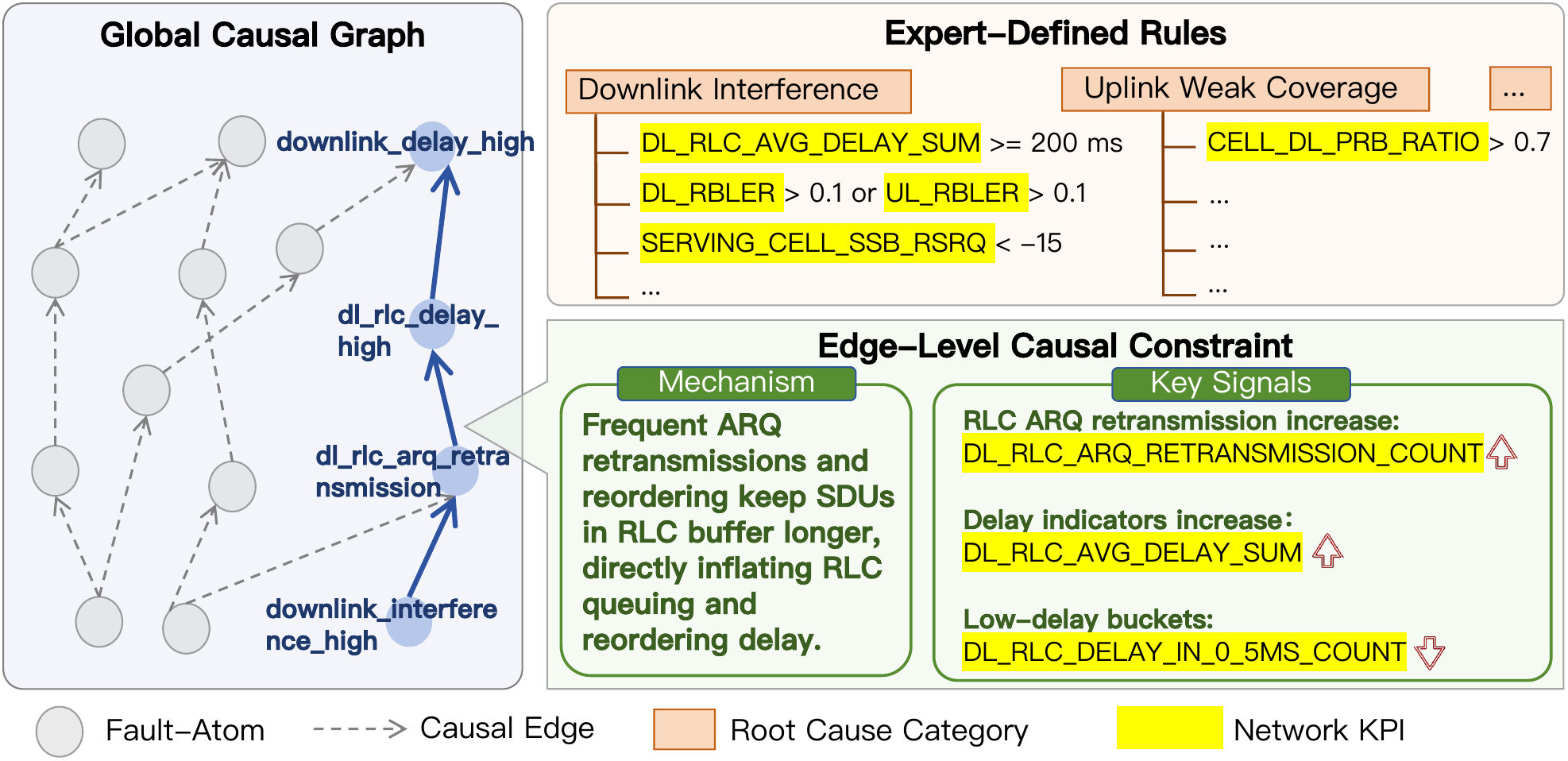}
    \caption{Illustration of the key components of the Knowledge Base.}
    \label{fig:kb_example}
\end{figure}

The \emph{Knowledge Base (KB)} formalizes diagnostic knowledge as a set of enforceable structural domain constraints.

\paragraph{Hierarchical Fault-Atom Abstraction:} Diagnosis is formulated over minimal semantic units, termed \emph{fault atoms}, which are organized into a three-layer hierarchy, as shown in the left panel of Fig.~\ref{fig:kb_example}: \emph{symptom} (e.g., \texttt{downlink\_delay\_high}), \emph{mechanism} (e.g., \texttt{dl\_rlc\_delay\_high}) and \emph{root cause} (e.g., \texttt{downlink\_interference\_high}). This hierarchy abstraction provides a semantically structured interface for controllable causal chain construction and verification.

\paragraph{Global Causal Graph:}
The KB maintains a global causal graph over fault atoms to encode valid cross-layer transitions and protocol-consistent fault propagation patterns. As illustrated in the left panel of Fig.~\ref{fig:kb_example}, one path captures that downlink interference evolves into retransmission abnormalities, intermediate delay buildup, and eventually high downlink delay. Candidate fault causal chains must follow graph-consistent connectivity, thereby restricting reasoning to structurally valid paths.

\paragraph{Edge-Level Causal Constraint:}
Each valid atom transition is annotated with domain knowledge describing its expected KPI-level manifestations, such as characteristic trends, symptomatic patterns, and underlying mechanisms, as illustrated in the bottom panel of Fig.~\ref{fig:kb_example}. During reasoning, the verifier uses these constraints to determine whether the observed KPI behaviors are causally consistent with the transition.

\paragraph{Expert-Defined Rules:}
The KB integrates interpretable rules based on operational thresholds, as shown in the top-right panel of Fig.~\ref{fig:kb_example}. Within the agent framework, these deterministic priors complement reasoning by providing auxiliary signals at key stages of diagnosis.

Overall, the KB suppresses spurious hypotheses and enhances interpretability without increasing model complexity, providing a lightweight yet effective way to impose domain-consistent structure on the diagnostic process.

\subsubsection{Historical Bank: Case-Based Reference}
\label{subsec:historical-bank}
The \emph{Historical Bank (HB)} is motivated by the observation that mobile network faults often recur through similar cross-layer causal structures, even though their raw KPI patterns may differ substantially across instances and are difficult to characterize with fixed thresholds~\cite{nouioua2021survey,10427016}. As a result, validated historical cases provide useful structural priors that complement the limited domain specificity of general-purpose LLMs. Each HB entry therefore pairs a representative KPI segment with its validated fault causal chain, serving as a case-based reference for reasoning.

\paragraph{Similarity-Guided Hypothesis Initialization:}
Retrieval in HB is performed in a learned representation space aligned with root-cause semantics. Specifically, the encoder of the pre-trained RCA model (\S\ref{subsec:tp}) projects multivariate KPI time series into a latent embedding space, in which samples with similar underlying causes are expected to lie closer. Feature-based retrieval over HB then identifies structurally similar historical cases. Rather than serving as direct answers, retrieved cases act as hypothesis initializers that provide candidate fault causal chain templates. 
This similarity-guided initialization transforms fault causal chain construction from open-ended combinatorial graph exploration into a constrained hypothesis verification process, significantly reducing the search space induced by the graph prior.

\paragraph{Evidence-Constrained Revalidation and Bias Control:}
Historical guidance from HB is always subject to revalidation under the current KPI observations and KB constraints before being accepted. This tight coupling between retrieval and verification confines HB to proposing structurally plausible hypotheses. As a result, \textit{QoEReasoner} uses historical priors to make reasoning more targeted and sample-efficient, while preserving consistency with present evidence.

\section{Experiment Setup}

\subsection{Implementation}
We implement \textit{QoEReasoner} with LangGraph~\cite{langgraph2024} under a ReAct-style control loop~\cite{yao2022react}. All tools are exposed as callable nodes with structured inputs and outputs, enabling reproducible orchestration and end-to-end traceability. To support reliable numeric perception, the \emph{Tool Pool} includes preprocessing utilities and lightweight CNN-based classifiers. Detailed architectures, training settings, and hardware configuration are provided in Appendix~\ref{appen:Implementation Details}.

\subsection{Datasets}
We evaluate \textit{QoEReasoner} on a real-world dataset from an operational mobile network. The dataset covers six root-cause categories and 21 expert-validated fault-chain templates, each capturing a propagation path from root cause to QoE degradation. It contains 300 sessions, including 130 abnormal sessions with expert-verified root-cause chains; the rest are normal sessions used for anomaly detection. \footnote{Detailed label definitions and the FCR vocabulary are provided in Appendix~\ref{appen:rcc_labels} and Appendix~\ref{appen:fault_chain}.} The limited scale reflects the high cost of industrial annotation, as each chain must be manually verified from multi-layer KPI traces by experienced engineers.
\section{Evaluation}

To provide a comprehensive evaluation of \textit{QoEReasoner} from both quantitative and qualitative perspectives, we first assess fault diagnosis performance against representative baselines across multiple paradigms. We then conduct module-level analyses to examine the effectiveness of key components within \textit{QoEReasoner}, followed by ablation studies on the \emph{Historical Bank (HB)} and \emph{Knowledge Base (KB)}. Finally, we present case studies and user evaluation to demonstrate practical utility and report quality.

\subsection{Metrics}

As defined in \S\ref{sec:task formu}, our framework addresses two core diagnostic tasks, AD and FCR, along with a derived evaluation dimension (root-cause classification), each evaluated with different metrics. We briefly summarize them below and defer the formal definitions to Appendix~\ref{appen:metrics}.

\begin{itemize}[itemsep=0em, topsep=0em, leftmargin=1em]
    \item \textbf{AD.} We report Precision, Recall, false alarm rate (FAR), and F1 score from the binary confusion matrix. These metrics respectively capture anomaly precision, detection coverage, false-alarm tendency on normal samples, and the balance between Precision and Recall.
    \item \textbf{FCR.} We report node-level precision/recall (node\_p, node\_r), edge precision (edge\_p), and exact chain match (chain\_em), averaged over samples with non-empty ground-truth chains. They evaluate node correctness, edge correctness, and exact recovery of the full fault chain. For the module-level analysis in \S\ref{exp:FCR module}, we additionally use several ranking metrics.\footnote{Formal definitions are provided in Appendix~\ref{FCR_module_metrics}.}
    \item \textbf{RCC.} We report top-1 accuracy (top1\_acc), macro-F1, and balanced accuracy (balanced\_acc). These reflect overall classification performance as well as class-balanced performance under label imbalance.
\end{itemize}

\subsection{Baseline Comparison}
\begin{table}
  \centering
  \caption{
  Anomaly detection (AD) performance comparison.
  ``--'' indicates the approach does not support this task.
  }
  \label{tab:ad_only}
    \resizebox{\columnwidth}{!}{%
    \setlength{\tabcolsep}{2.5pt}
    \fontsize{9pt}{11pt}\selectfont
    \begin{tabular}{l l c c c c}
      \toprule
      \textbf{Method} & \textbf{Paradigm}
      & \textbf{Precision $\uparrow$} & \textbf{Recall $\uparrow$} & \textbf{FAR $\downarrow$} & \textbf{F1 $\uparrow$} \\
      \midrule

      Rule & Heuristic
      & 0.6833 & 0.6496 & 0.2883 & 0.6520 \\

      TSTCC~\cite{eldele2021ts2vec} & DL
      & -- & -- & -- & -- \\

      CATCC~\cite{eldele2023self} & DL
      & -- & -- & -- & -- \\

      \midrule
      DeepSeek-V3.2 & LLM
      & 0.2452 & \textbf{1.0000} & 1.0000 & 0.3939 \\

      + RCA-Agent~\cite{xu2025openrca} & Agent
      & 0.4567 & \textbf{1.0000} & 1.0000 & 0.6270 \\

      + \textit{\textbf{QoEReasoner}} & Agent
      & \textbf{0.8600} & 0.8686 & \textbf{0.1472} & \textbf{0.8500} \\

      \midrule
      GPT-5.2 & LLM
      & 0.3643 & 0.6623 & 0.3755 & 0.4700 \\

      + RCA-Agent~\cite{xu2025openrca} & Agent
      & 0.5146 & 0.6423 & 0.5092 & 0.5714 \\

      + \textit{\textbf{QoEReasoner}} & Agent
      & \textbf{0.8600} & \textbf{0.8686} & \textbf{0.1472} & \textbf{0.8500} \\

      \midrule
      Qwen3-14B & LLM
      & 0.2637 & \textbf{0.9863} & 0.9095 & 0.4162 \\

      + RCA-Agent~\cite{xu2025openrca} & Agent
      & 0.5000 & 0.5693 & 0.4785 & 0.5324 \\

      + \textit{\textbf{QoEReasoner}} & Agent
      & \textbf{0.8567} & 0.8686 & \textbf{0.1534} & \textbf{0.8479} \\

      \bottomrule
    \end{tabular}
    }
\end{table}

\begin{table*}
  \centering
  \caption{
Comparison of structured reasoning (FCR) and root-cause classification (RCC) performance. Node/edge metrics evaluate the structural correctness of inferred fault chains, and the root-cause category is derived from the first node of each chain.
  }
  \label{tab:fcr_rc}
    \resizebox{\textwidth}{!}{%
    \setlength{\tabcolsep}{6pt}
    \fontsize{10pt}{11pt}\selectfont
    \begin{tabular}{l l
    c c c c |
    c c c}
      \toprule
      \textbf{Method} & \textbf{Paradigm}
      & \multicolumn{4}{c|}{\textbf{Reasoning (FCR)}}
      & \multicolumn{3}{c}{\textbf{Root-Cause (RCC)}} \\
      \cmidrule(lr){3-6} \cmidrule(lr){7-9}
      &
      & node\_p $\uparrow$ & node\_r $\uparrow$ & edge\_p $\uparrow$ & chain\_em $\uparrow$
      & top1\_acc $\uparrow$ & macro\_F1 $\uparrow$ & balanced\_acc $\uparrow$ \\
      \midrule

      Rule & Heuristic
      & -- & -- & -- & --
      & 0.0571 & 0.1800 & 0.2444 \\

      TSTCC~\cite{eldele2021ts2vec} & DL
      & -- & -- & -- & --
      & 0.3143 & 0.0965 & 0.1073 \\

      CATCC~\cite{eldele2023self} & DL
      & -- & -- & -- & --
      & 0.2857 & 0.0901 & 0.0976 \\

      \midrule
      DeepSeek-V3.2 & LLM
      & 0.1488 & 0.1036 & 0.0048 & 0.0000
      & 0.0286 & 0.0870 & 0.0098 \\

      + RCA-Agent~\cite{xu2025openrca} & Agent
      & 0.1126 &  0.1071 & 0.0214 & 0.0000
      & 0.0143 & 0.0211 & 0.2000 \\

      + \textit{\textbf{QoEReasoner}} & Agent
      & \textbf{0.4271} & \textbf{0.4264} & \textbf{0.3095} & \textbf{0.2857}
      & \textbf{0.5714} & \textbf{0.5333} & \textbf{0.5518} \\



      \midrule
      GPT-5.2 & LLM
      & 0.3083 & 0.2207 & 0.0476 & 0.0000
      & 0.3623 & \textbf{0.6757} & 0.1220 \\

      + RCA-Agent~\cite{xu2025openrca} & Agent
      & 0.1179 & 0.0893 & 0.0000 & 0.0000
      & 0.1714 & 0.1555 & 0.3160 \\

      + \textit{\textbf{QoEReasoner}} & Agent
      & \textbf{0.5414} & \textbf{0.5400} & \textbf{0.4286} & \textbf{0.4000}
      & \textbf{0.6714} & 0.6171 & \textbf{0.5859} \\

      \midrule
      Qwen3-14B & LLM
      & 0.1738 & 0.0879 & 0.0143 & 0.0000
      & 0.0000 & 0.0000 & 0.0000 \\

      + RCA-Agent~\cite{xu2025openrca} & Agent
      & 0.0048 & 0.0036 & 0.0000 & 0.0000
      & 0.0143 & 0.0069 & 0.2000 \\

      + \textit{\textbf{QoEReasoner}} & Agent
      & \textbf{0.4464} & \textbf{0.4429} & \textbf{0.3286} & \textbf{0.3000}
      & \textbf{0.6000} & \textbf{0.5829} & \textbf{0.5566} \\

      \bottomrule
    \end{tabular}
    }
\end{table*}

\paragraph{Baselines.}
We compare \textit{QoEReasoner} against representative baselines from rule-based, deep learning, pure LLM, and agent-based paradigms. Specifically, for the DL paradigm, we consider TSTCC~\cite{eldele2021ts2vec} and CATCC~\cite{eldele2023self}; for the agent paradigm, we use RCA-Agent~\cite{xu2025openrca}. More details are provided in Appendix~\ref{appen:baselines}.

\paragraph{Multi-Task Performance Advantage.}
From the perspective of task coverage, the compared baselines exhibit clear limitations across the diagnosis tasks. Rule-based heuristics and conventional deep learning models can deliver acceptable performance on anomaly detection, but they remain weak at root-cause identification and do not support fault causal chain analysis. Pure LLMs, while capable of generating free-form reasoning traces, fail to produce stable and structurally valid fault chains, and their decision quality varies considerably across different model backbones. RCA-Agent improves anomaly detection over pure LLMs, largely due to its controller-executor workflow, yet its lack of task-specific domain constraints and customized diagnostic knowledge makes it ineffective for more demanding reasoning tasks such as fault causal reasoning. In contrast, \textit{QoEReasoner} consistently supports anomaly detection, fault causal reasoning, and root-cause identification within a unified framework, demonstrating a level of cross-task compatibility that prior approaches inherently lack. 
Additionally, these results collectively demonstrate that the framework fulfills the requirements of causally grounded fault-chain reasoning (R2) and coherent cross-task coordination (R3).

From a performance standpoint, \textit{QoEReasoner} further delivers clear gains across all tasks. In anomaly detection, its structured tool invocation and verification pipeline improve reliability by suppressing oversensitive judgments and reducing spurious predictions. At the reasoning and decision layers, the advantage becomes even more evident: pure LLMs struggle with node- and edge-level structural correctness, while deep learning models, although sometimes competitive for flat classification, lack explicit reasoning capability and therefore cannot construct interpretable causal structures. By coordinating structured evidence, protocol-aware constraints, and task-oriented reasoning modules within the agent loop, \textit{QoEReasoner} not only improves the structural validity of predicted fault chains but also achieves stronger root-cause classification accuracy. Overall, these results show that the superiority of \textit{QoEReasoner} comes not merely from using an LLM-based agent architecture, but from embedding structured evidence coordination and constraint-aware reasoning into a unified multi-task diagnostic framework.

\paragraph{Robustness Across LLM Scales.}
Another important finding is that the performance gains of \textit{QoEReasoner} remain consistent across different LLM backbones. While pure LLM baselines show substantial variation in standalone reasoning and classification quality, their agent-enhanced counterparts achieve much more stable and uniformly strong performance. This suggests that the effectiveness of \textit{QoEReasoner} can compensate for backbone differences through structured tool support, constrained reasoning, and evidence verification, enabling even smaller LLMs to attain performance competitive with much larger models. Such backbone-agnostic robustness is particularly valuable for practical deployment, as it improves model flexibility while reducing dependence on expensive high-end LLMs.

\subsection{Module Analysis}

This section evaluates the internal design of \textit{QoEReasoner}, including planner-guided anomaly perception and the FCR module for structured fault causal chain reasoning.

\subsubsection{Anomaly Perception under Planner Analysis}
\label{subsubsec:planner-analysis}

\begin{figure}
  \centering
  \includegraphics[width=\linewidth]{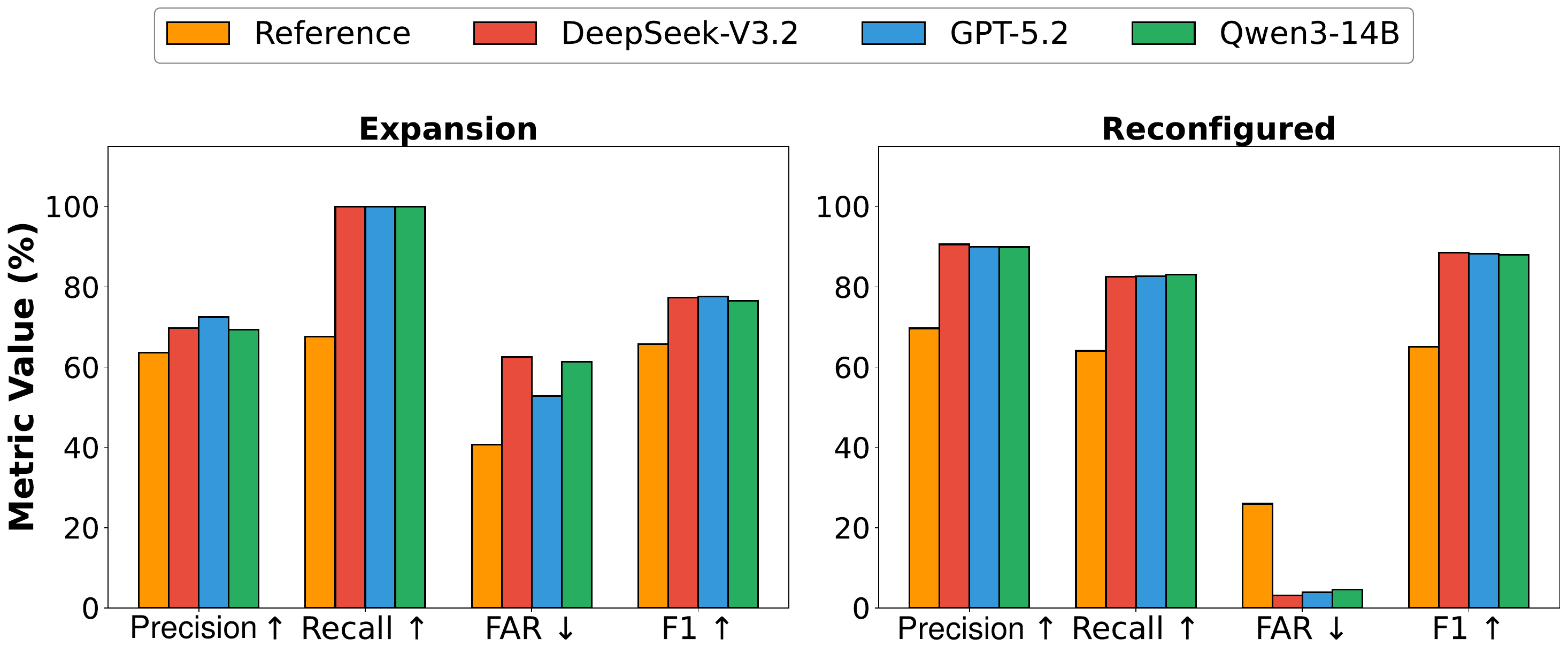}
  \caption{
   Performance of QoE degradation detection under different planner-induced structural deviation categories.
  }
  \label{fig:planner_path_dev}
\end{figure}

To examine anomaly detection under planner-guided execution, we use a rule-based AD $\rightarrow$ FCR workflow as a conservative deterministic baseline. Planner-generated paths are then compared against this reference in terms of structural deviation and AD performance, where \textit{Expansion} extends the path with additional evidence and \textit{Reconfigured} reorganizes the reasoning structure.

Across LLM backbones, about 20\% of cases fall into \textit{Expansion}, while nearly 80\% are \textit{Reconfigured}. As shown in Fig.~\ref{fig:planner_path_dev}, \textit{Expansion} slightly increases false alarms, whereas \textit{Reconfigured} improves overall AD performance by reducing false alarms without sacrificing recall. This suggests that restructuring diagnostic dependencies yields more reliable perception (R1). The results are consistent with the planner’s design. It treats the rule as a structural prior and adaptively adds tools or reordering steps when beneficial. Together, these findings support that stateful planning enables coherent cross-task reasoning (R3).

\subsubsection{Fault Causal Chain Reasoning Module}
\label{exp:FCR module}

\begin{table}
\centering
  \caption{Stage-wise ranking of the ground-truth chain among FCR candidate chains with GPT-5.2 as the verifier backbone. MRR and Top-5 AUC characterize the concentration of the ranking toward the ground-truth chain, while R@1/3/5 measure its cumulative recall under different cutoffs.}
\label{tab:fcr_stage_analysis}
\resizebox{\columnwidth}{!}{%
\setlength{\tabcolsep}{5pt}
\fontsize{9pt}{11pt}\selectfont
\begin{tabular}{lccccc}
\toprule
\textbf{Stage} & \textbf{MRR} $\uparrow$ & \textbf{AUC} $\uparrow$ & \textbf{R@1} $\uparrow$ & \textbf{R@3} $\uparrow$ & \textbf{R@5} $\uparrow$ \\
\midrule
Retrieval & 0.6003 & 3.1714 & 0.3429 & 0.6857 & 0.8714 \\
Verifier  & 0.5817 & 3.4143 & 0.3426 & 0.7429 & 0.9000 \\
Adjuster  & 0.6206 & 3.5714 & 0.4000 & 0.7571 & 0.9142 \\
\bottomrule
\end{tabular}
}
\end{table}

Table~\ref{tab:fcr_stage_analysis} highlights the complementary roles of the three stages in FCR. At retrieval, the ground-truth (GT) chain already achieves reasonably strong coverage in the candidate set, indicating that similarity-guided retrieval can form a concentrated hypothesis space. However, the ranking is still coarse, as shown by only moderate MRR and limited Recall@1. Verification and adjustment then further improve MRR and Recall@1, showing that score fusion and consistency-aware reweighting effectively promote the GT chain toward the top of the ranking. This confirms that candidate pruning and ranking refinement play distinct yet complementary roles.
Overall, the stage-wise trend validates the necessity of the multi-stage FCR design. Retrieval ensures coverage, verification enforces causal consistency, and adjustment sharpens ranking concentration, together enabling stable and interpretable fault causal chain reconstruction.

\begin{figure}
    \centering
    \includegraphics[width=\linewidth]{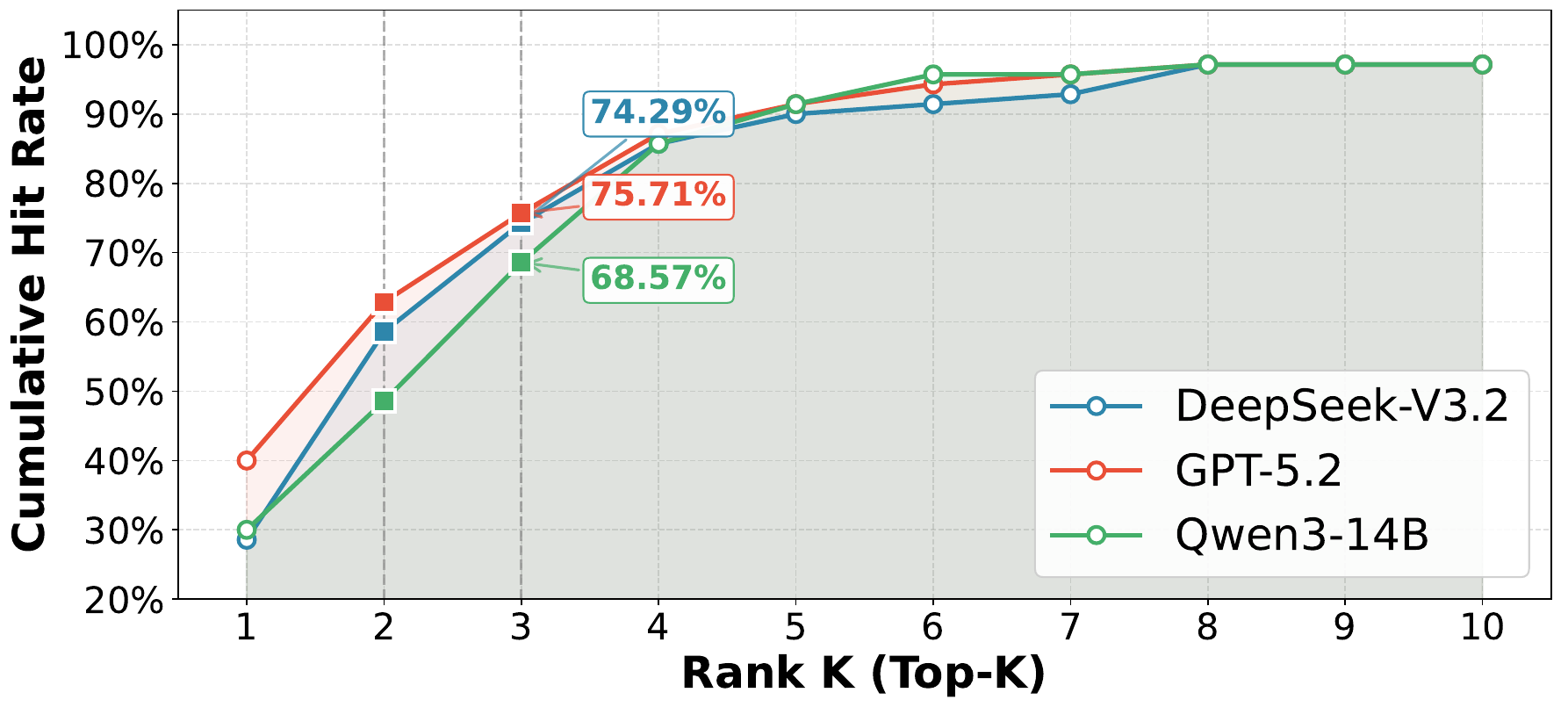}
    \caption{The cumulative Top-$K$ hit rate of the FCR module.}
    \label{fig:FCR_cumulative_hit_rate}
\end{figure}

Notably, FCR remains robust across different LLM backbones, enabling lightweight models such as Qwen-14B to achieve performance comparable to larger models such as GPT-5.2 within the same pipeline. Although different LLMs exhibit markedly different pruning behaviors at the verifier stage, the final GT-chain ranking after adjustment remains consistently strong. Fig.~\ref{fig:FCR_cumulative_hit_rate} further supports this finding by showing closely aligned Top-$K$ hit curves across backbones, where correct hypotheses are progressively concentrated into compact candidate sets and the Top-3 predictions already cover the GT chain in most cases. These results further confirm the effectiveness of \textit{QoEReasoner} in concentrating correct fault-chain hypotheses through structured reasoning.

\subsection{Ablation Study}
\label{sec:ablation}

\begin{figure}
\centering
\includegraphics[width=\columnwidth]{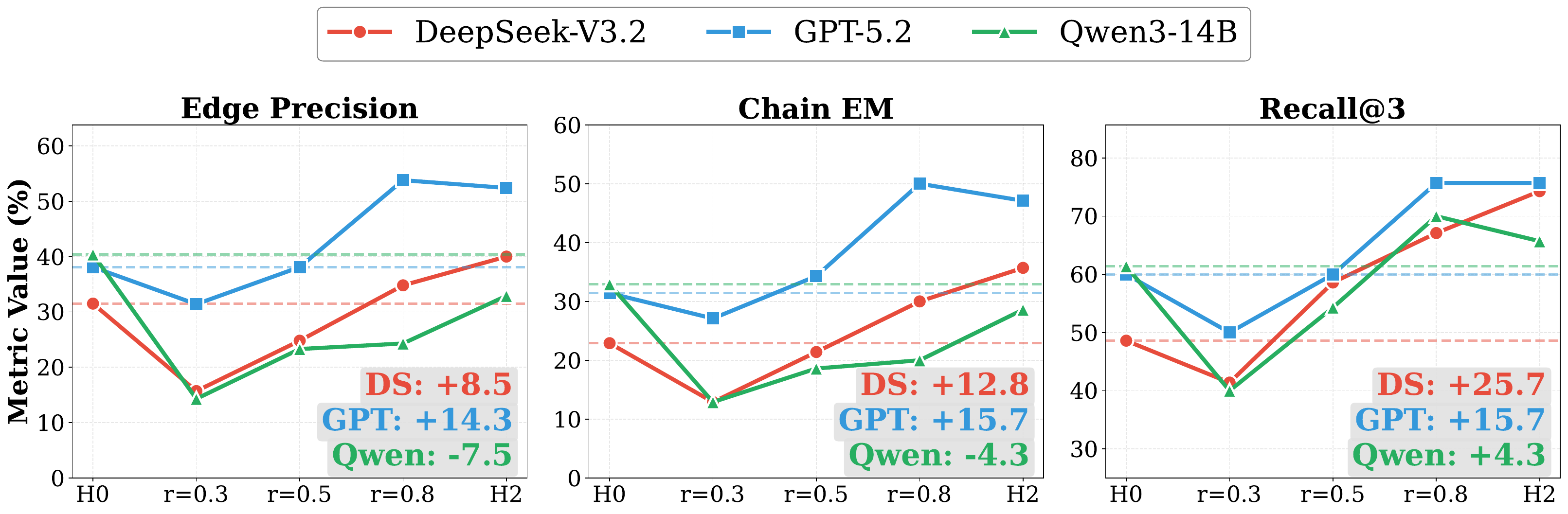}
\caption{Impact of HB coverage on candidate-level fault causal chain quality. }
\label{fig:hb_candidate_space}
\end{figure}

To assess the individual and joint effects of KB and HB in \textit{QoEReasoner}, we perform ablation studies across multiple regimes, as summarized in Fig.~\ref{fig:hb_candidate_space} and Table~\ref{tab:ablation_fcr_only}.

\paragraph{Complementary structural roles of KB and HB.}
KB and HB improve FCR through complementary mechanisms. Enabling KB alone substantially improves structural validity, as reflected in stronger edge- and chain-level performance, because protocol-aware causal constraints explicitly enforce feasible cross-layer propagation paths. However, without HB, performance remains limited by the lack of data-driven priors, indicating that structural correctness alone is insufficient to concentrate hypotheses around semantically aligned root causes. Another observation is that HB alone still yields clearly weaker structural quality than the full model, since historical templates cannot by themselves exclude structurally invalid transitions. Overall, HB reshapes the hypothesis space with distributional priors, while KB defines its feasible boundaries.

\paragraph{Impact of HB coverage and prior bias.}

We further observe that insufficient HB coverage can significantly degrade performance. Similar trends are consistently observed across different LLM backbones in Fig.~\ref{fig:hb_candidate_space}, where limited historical retrieval tends to introduce biased priors that restrict exploration without providing sufficient structural diversity. Nevertheless, even under such low-coverage regimes, the KB-only baseline remains competitive on several metrics, suggesting that structurally grounded reasoning enforced by KB can be more reliable than weak or noisy priors induced by sparse historical templates.

Overall, these results show that \textit{QoEReasoner} improves reasoning effectiveness by grounding diagnosis in structured constraints and historical reference.

\begin{table}
\centering
\caption{End-to-End Fault Causal Chain Quality under KB and HB Ablation (GPT-5.2).}
\label{tab:ablation_fcr_only}
\resizebox{\columnwidth}{!}{%
\setlength{\tabcolsep}{2.5pt}
\fontsize{9pt}{11pt}\selectfont
\begin{tabular}{c c c c c c}
\toprule
\multirow{2}{*}{\textbf{HB}} 
& \multirow{2}{*}{\textbf{KB}} 
& \multicolumn{4}{c}{\textbf{Fault Causal Chain Quality (FCR)}} \\
\cmidrule(lr){3-6}
& 
& \textbf{node\_p} $\uparrow$ 
& \textbf{node\_r} $\uparrow$ 
& \textbf{edge\_p} $\uparrow$ 
& \textbf{chain\_em} $\uparrow$ \\

\midrule
\xmark & \xmark 
& 0.3083 & 0.2207 & 0.0476 & 0.0000 \\

\xmark & \cmark  
& 0.4957 & 0.5050 & 0.3810 & 0.3143 \\

\pmark ($r=0.5$) & \cmark 
& 0.4550 & 0.4514 & 0.3476 & 0.3143 \\

\cmark & \xmark 
& 0.4286 & 0.4257 & 0.2762 & 0.2429 \\

\cmark & \cmark 
& \textbf{0.5414} 
& \textbf{0.5400} 
& \textbf{0.4286} 
& \textbf{0.4000} \\
\bottomrule
\end{tabular}
}
\end{table}

\subsection{Case Study and Expert Evaluation}

\begin{figure*}
    \centering
    \begin{subfigure}[t]{0.40\textwidth}
        \centering
        \includegraphics[width=\linewidth]{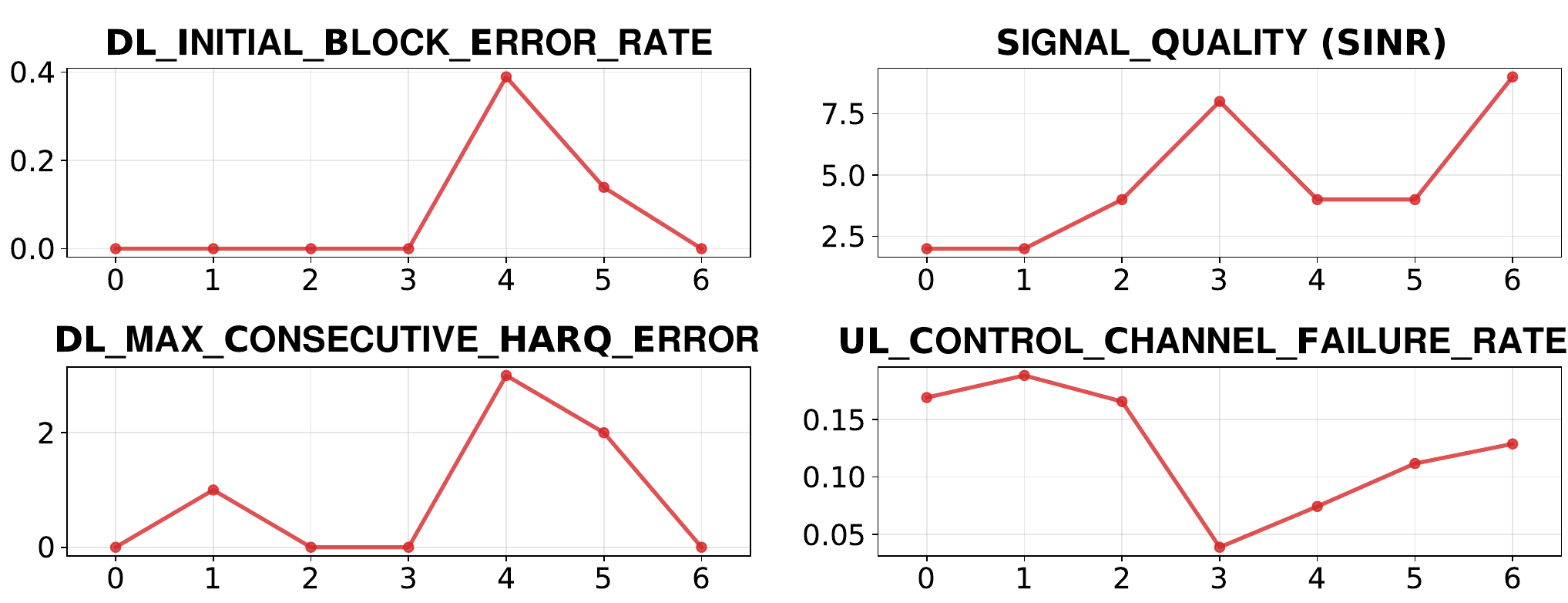}
        \caption{Temporal dynamics of partial KPIs.}
        \label{fig:case_kpi}
    \end{subfigure}
    \hfill
    \begin{subfigure}[t]{0.41\textwidth}
        \centering
        \includegraphics[width=\linewidth]{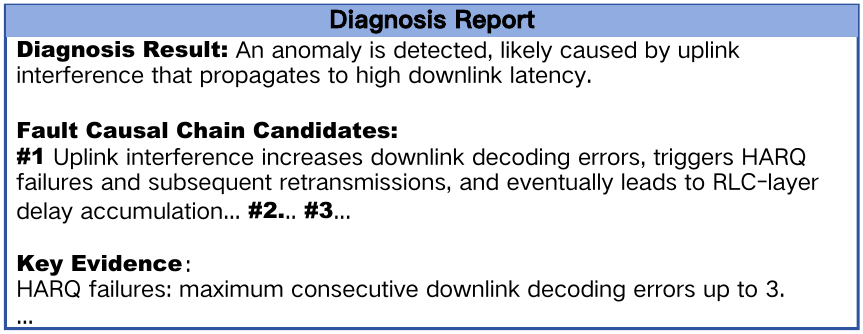}
        \caption{Illustrative diagnostic report.}
        \label{fig:case_report}
    \end{subfigure}
    \hfill
    \begin{subfigure}[t]{0.17\textwidth}
        \centering
        \includegraphics[width=\linewidth]{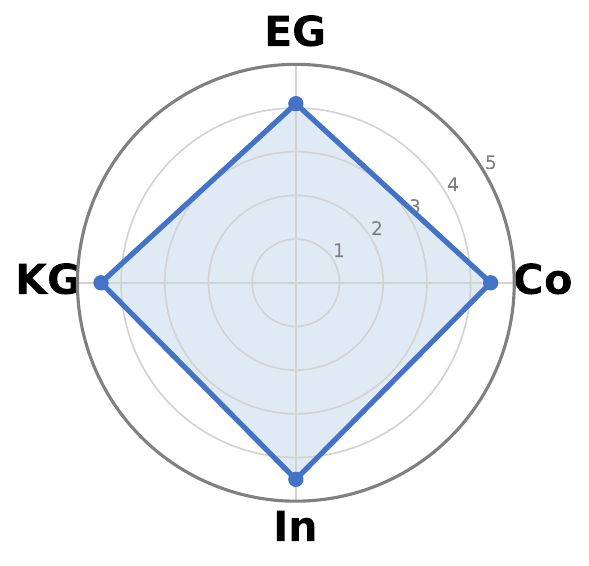}
        \caption{Subjective Expert User study.}
        \label{fig:user_study_subjective}
    \end{subfigure}
    \caption{Case study and expert subjective evaluation of \textit{QoEReasoner}.}
    \label{fig:case_user_study}
\end{figure*}

We present a case study and expert evaluation to qualitatively assess the interpretability and practical utility of \textit{QoEReasoner}.

\paragraph{Case Study.}
As shown in Fig.~\ref{fig:case_user_study}, \textit{QoEReasoner} converts complex multi-layer KPI dynamics into a concise and evidence-grounded diagnostic report. In this example, the system identifies \emph{uplink interference} as the primary cause. The generated report contains the final diagnosis, the highest-ranked fault chain together with alternative hypotheses, and a natural-language explanation that connects observed KPI abnormalities with the inferred propagation path, thereby better matching expert diagnostic practice.
We further quantify the inference overhead of this case. The end-to-end diagnosis uses 166{,}375 tokens, comprising 117{,}888 cached input tokens, 35{,}929 non-cached input tokens, and 12{,}558 output tokens, for a total API cost of about \$0.02 when the retrieval hyperparameter in \S\ref{subsec:rc-reasoning} is set to $n=5$. This result indicates that \textit{QoEReasoner} can deliver interpretable and operationally meaningful diagnosis at low per-session monetary cost.

\paragraph{Subjective Measurement.}
We conducted a user study in which three domain experts jointly reviewed 50 diagnostic reports generated by GPT-5.2 + \textit{QoEReasoner}. Each report was rated on four dimensions, namely Correctness (Co), Evidence Grounding (EG), Knowledge Grounding (KG), and Interpretability (In), using a five-point Likert scale.\footnote{The detailed scoring rubric is provided in Appendix~\ref{appen:user-study}.} As shown in Fig.~\ref{fig:user_study_subjective}, all average scores exceed 4, suggesting that the reports are generally perceived as accurate, knowledge-grounded, and easy to interpret. Interpretability achieves the highest score, while the slightly lower Evidence Grounding score suggests room to further strengthen the linkage between conclusions and supporting KPI observations. In addition, \textit{QoEReasoner} typically completes one diagnostic session in about 3 minutes, whereas manual expert analysis often takes 30--60 minutes, demonstrating its potential to improve diagnostic efficiency with interpretable and operationally meaningful outputs.

\begin{figure}
    \centering
    \includegraphics[width=1\linewidth]{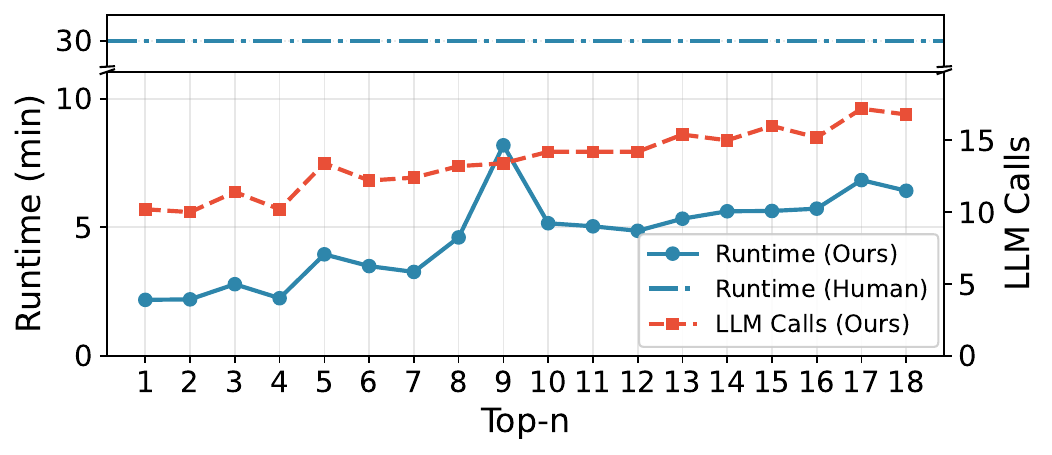}
    \caption{Runtime and LLM API calls versus the Top-$n$ retrieval budget.}
    \label{fig:efficiency_combined_vs_topn}
\end{figure}

\section{Discussion}
\label{sec:discussion}

\paragraph{Runtime Cost and Scaling.}
\textit{QoEReasoner} demonstrates practically acceptable efficiency together with predictable scaling behavior. As shown in Fig.~\ref{fig:efficiency_combined_vs_topn}, diagnosis latency increases with the Top-$n$ retrieval budget and closely tracks the number of LLM invocations, indicating that runtime is primarily dominated by multi-round planning and causal verification. Empirically, we observe that \textit{QoEReasoner} can already produce reports with expert-level quality within about three minutes per session. This runtime is substantially more efficient than manual expert analysis, which typically requires about 30--60 minutes. Moving forward, \textit{QoEReasoner} can be further accelerated by more efficient techniques such as local/lightweight LLM serving and parallel verifier execution.


\paragraph{Deployment and Extensibility.}
From a deployment standpoint, \textit{QoEReasoner} remains lightweight because it reuses pre-trained offline models and external LLM services without requiring task-specific fine-tuning at inference time. More importantly, through a modular design paradigm in which the framework is decoupled through explicit interfaces, new tools or domain-specific components can be integrated without redesigning the overall workflow. This property is particularly valuable in RANs environments, where KPI definitions, network conditions, and operational practices evolve continuously. Although post-diagnosis optimization is outside the scope of this paper, the same modular design may support future integration of components such as simulation-backed validation or recommendation modules, thereby enabling broader detect--diagnose--optimize workflows.

\paragraph{Beyond Single-Chain Diagnosis.}
Our current formulation adopts a practical single-dominant-chain assumption to enable tractable structured diagnosis. In real RANs environments, however, QoE degradation may arise from multiple concurrent and interacting fault propagation paths rather than from a single isolated chain. For instance, the same session may simultaneously reflect retransmission-related and control-channel-related symptoms triggered by a shared upstream condition. Such ambiguity is difficult even for human experts to fully enumerate and also limits the completeness of available supervision. For this reason, \textit{QoEReasoner} is designed to return multiple plausible fault chains, making the framework more useful when degradation processes are only partially observable or involve concurrent causes, while also providing a natural foundation for future extensions toward structured multi-chain diagnosis.

\paragraph{Toward Adaptive Agentic Diagnosis.}
The present framework already offers a structured diagnosis pipeline, which provides a basis for improvement without changing the overall architecture. In particular, the planner could gradually become more adaptive by learning how to balance case complexity and historical diagnostic outcomes when selecting subsequent actions. In this way, \textit{QoEReasoner} may evolve toward a more experience-driven agentic diagnosis framework, offering a practical path toward AI-native RANs diagnosis.

\section{Related Work}
\paragraph{Anomaly Diagnosis in RANs.}
Traditional RANs diagnosis relies on expert rules~\cite{kane2015system}, but suffers from limited scalability. Data-driven methods including machine learning~\cite{sundqvist2020boosted, dimopoulos2015identifying, chen2019machine} and deep learning~\cite{fida2023bottleneck} have become dominant but lack explicit fault causal chain construction or causal verification. These approaches treat perception, reasoning, and decision-making as loosely coupled components without explicit diagnostic state or constraint-aware verification.
In contrast, \textit{QoEReasoner} integrates perception, planning, and reasoning within a unified loop, maintaining explicit states and producing traceable fault causal chains.

\paragraph{LLM-Enabled Agentic AI in Wireless Networks.}
Recent advances have stimulated interest in LLM integration into wireless systems~\cite{boateng2025survey, jiang2024large}, from natural language interaction for intent management~\cite{mekrache2024intent, tong2024connectgpt} to agent-based paradigms characterized by perception–reasoning–action loops and explicit state tracking~\cite{zhang2026toward}.
Agentic frameworks have been explored for mobile edge intelligence, semantic communication, and network optimization~\cite{qu2025mobile, luo2025toward, sun2025edge, tong2025wirelessagent}, with architectural visions incorporating agents into 6G and O-RANs ecosystems~\cite{li2024agent, salama2025edge, jiang2026large, lin2025pushing}.
However, agentic frameworks for causal root-cause reasoning in RANs diagnosis remain under-explored, given unique challenges of cross-layer coordination and strict domain constraints. \textit{QoEReasoner} introduces a constraint-aware diagnostic loop with deterministic perception and verifiable fault causal chain reasoning.

\section{Conclusion}
We propose \textit{QoEReasoner}, a unified agentic framework for automated and explainable QoE diagnosis in RANs. Our framework features three key technical designs: tool-grounded deterministic perception modules for reliable numeric analysis, a Knowledge Base and Historical Bank that provide domain-consistent structural constraints and data-driven priors throughout the diagnostic process, and a stateful planner and reporter that coordinate the multi-task diagnostic loop and synthesize interpretable outputs. Together, these components enable \textit{QoEReasoner} to couple trustworthy KPI understanding with protocol-consistent causal reasoning under a unified diagnostic workflow. This design allows the system to support end-to-end diagnosis across anomaly perception, reasoning, and report generation. Our experiments on real-world operational data demonstrate consistent improvements over baselines from several prior paradigms, with robust performance across diverse LLM backbones and validated interpretability from domain expert evaluation. These results highlight the strong potential of combining deterministic tools, structured domain knowledge, and LLM-based coordination in a unified diagnostic loop to advance scalable network diagnosis.



\bibliographystyle{ACM-Reference-Format}
\bibliography{ref}

\appendix

\section{Preliminary LLM Study Setup}\label{appen:prelim_llm_setup}
This appendix summarizes the experimental setup for the preliminary study in \S\ref{subsec:preliminary}.
We consider prompt-based anomaly detection on cellular KPI sequences and evaluate representative, recently released industry-leading LLM families, including DeepSeek-V3.2, Gemini series, GPT series, and Qwen3 family~\cite{liu2025deepseek, comanici2025gemini, hurst2024gpt, yang2025qwen3}.
The evaluation uses 350 labeled KPI sequence samples, consisting of 175 normal and 175 anomalous cases.

\section{Implementation Details}
\label{appen:Implementation Details}
We implement \textit{QoEReasoner} with LangGraph~\cite{langgraph2024} and organize the workflow under a ReAct-style control loop~\cite{yao2022react}. All tools are exposed as callable nodes with structured inputs and outputs, which facilitates reproducible orchestration and end-to-end traceability.

To provide reliable numeric perception, we implement data preprocessing utilities and lightweight CNN-based classifiers as deterministic tools. For anomaly detection (AD), we train a supervised binary classifier on labeled QoE session samples (\emph{normal} vs.\ \emph{anomalous}). Given a multivariate KPI sequence $\mathbf{X}\in\mathbb{R}^{T\times M}$, the model outputs an anomaly probability, which is used by the planner for confidence tracking and stopping decisions.

For root-cause classification (RCC), we train a supervised multi-class classifier on labeled anomalous samples with root-cause category annotations. The predicted category distribution is then used as candidate evidence, which can be further verified or corrected by the knowledge-constrained fault causal chain reasoning module.

All experiments are conducted on a single NVIDIA RTX 3090 GPU.

\section{Fault Chain Vocabulary}\label{appen:fault_chain}
We show the fault chain vocabulary in Table~\ref{tab:fcr_vocabulary}.

\begin{table}[t]
\centering
\caption{FCR vocabulary of anomaly phenomena.}
\label{tab:fcr_vocabulary}
\resizebox{\linewidth}{!}{%
\setlength{\tabcolsep}{1pt}
\fontsize{9pt}{10pt}\selectfont
\begin{tabular}{c}
\toprule
\textbf{Vocabulary Atom} \\
\midrule
\texttt{cell\_level\_high\_CCE\_load\_or\_allocation\_failure} \\
\texttt{user\_level\_high\_uplink\_CCE\_allocation\_failure} \\
\texttt{sr\_scheduling\_delay\_high} \\
\texttt{uplink\_delay\_high} \\
\texttt{uplink\_weak\_coverage} \\
\texttt{ul\_low\_spectral\_efficiency} \\
\texttt{ul\_rlc\_many\_segments} \\
\texttt{ul\_rlc\_delay\_high} \\
\texttt{uplink\_interference\_high} \\
\texttt{dl\_high\_error\_rate} \\
\texttt{dl\_arq\_high} \\
\texttt{dl\_rlc\_delay\_high} \\
\texttt{user\_level\_high\_downlink\_CCE\_allocation\_failure} \\
\texttt{dl\_rlc\_many\_segments} \\
\texttt{dl\_low\_small\_packet\_ratio} \\
\texttt{downlink\_delay\_high} \\
\texttt{ul\_air\_interface\_dtx\_rbler} \\
\texttt{ul\_rlc\_arq\_retransmission} \\
\texttt{downlink\_weak\_coverage} \\
\texttt{dl\_air\_interface\_dtx\_rbler} \\
\texttt{dl\_rlc\_arq\_retransmission} \\
\texttt{downlink\_interference\_high} \\
\texttt{cell\_downlink\_PRB\_utilization\_high} \\
\texttt{user\_high\_traffic\_demand} \\
\texttt{high\_CCE\_aggregation\_level} \\
\texttt{high\_CCE\_allocation\_failure} \\
\texttt{packet\_aggregation} \\
\texttt{rlc\_arq\_retransmission} \\
\texttt{dl\_spectral\_efficiency\_low} \\
\texttt{uplink\_rlc\_fragmentation\_high} \\
\bottomrule
\end{tabular}
}
\end{table}

\section{Root-Cause Category Labels}\label{appen:rcc_labels}

We show all root-cause category labels in Table~\ref{tab:category_description}.

\begin{table*}
\caption{Description of the Root-Cause Categories.}
\label{tab:category_description}
\centering
\resizebox{\textwidth}{!}{%
\setlength{\tabcolsep}{3pt}
\fontsize{8pt}{11pt}\selectfont
\begin{tabular}{|c|c|l|} 
\hline
\textbf{ID} & \textbf{Category} & \textbf{Description} \\
\hline
1 & Uplink Interference & External noise or frequency overlap causes decoding errors and uplink throughput degradation. \\
\hline
2 & Uplink Weak Coverage & Low uplink signal strength due to path loss or deep indoor scenarios, leading to increased latency. \\
\hline
3 & Downlink Interference & Strong interference from neighboring cells degrades SINR and reduces downlink data rates. \\
\hline
4 & Downlink Weak Coverage & Weak downlink signal at the cell edge or under obstruction leads to poor service quality. \\
\hline
5 & Traffic Channel Overload & Overload of data channels caused by excessive user traffic demand, resulting in congestion and increased latency. \\
\hline
6 & Control Channel Overload & Congestion on control channels delays resource scheduling and increases access failures. \\
\hline
\end{tabular}
}
\end{table*}

\section{Metrics}\label{appen:metrics}

\paragraph{Notation.}
Let $N$ denote the number of evaluated samples for AD or RCC after task-specific filtering.
For AD, let $(TP, FP, TN, FN)$ be the binary confusion counts.
For RCC, let $\mathcal{C}$ denote the set of ground-truth root-cause classes after filtering, and let $(y_i,\hat{y}_i)$ be the ground-truth and predicted class labels for sample $i$.
For FCR, let $g_i$ and $p_i$ denote the ground-truth and predicted fault chains for sample $i$, respectively, where each chain is represented as an ordered node sequence. Let $M$ be the number of samples with non-empty ground-truth chains.

\paragraph{AD metrics.}
We use standard binary classification metrics to evaluate anomaly detection.
Precision measures the correctness of predicted anomalies, Recall measures the coverage of true anomalies, FAR measures the false-alarm tendency on normal samples, and F1 balances Precision and Recall:
\begin{align}
\mathrm{Precision} &= \frac{TP}{TP+FP},\\
\mathrm{Recall} &= \frac{TP}{TP+FN},\\
\mathrm{FAR} &= \frac{FP}{FP+TN},\\
\mathrm{F1} &= \frac{2\cdot \mathrm{Precision}\cdot \mathrm{Recall}}{\mathrm{Precision}+\mathrm{Recall}}.
\end{align}

\paragraph{RCC metrics.}
For root-cause classification, top-1 accuracy measures overall prediction correctness, while macro-F1 and balanced accuracy emphasize class-balanced performance.

Top-1 accuracy is defined as
\begin{align}
\mathrm{top1\_acc}=\frac{1}{N}\sum_{i=1}^{N}\mathbf{1}[\hat{y}_i=y_i].
\end{align}

For each class $c\in\mathcal{C}$, let $TP_c$, $FP_c$, and $FN_c$ denote the one-vs-rest confusion counts. Then
\begin{align}
P_c &= \frac{TP_c}{TP_c+FP_c},\quad
R_c = \frac{TP_c}{TP_c+FN_c},\\
F1_c &=
\begin{cases}
\frac{2P_cR_c}{P_c+R_c}, & P_c+R_c>0,\\
0, & \text{otherwise}.
\end{cases}
\end{align}
Macro-F1 averages per-class F1 scores and thus treats all classes equally:
\begin{align}
\mathrm{macro\_F1} &= \frac{1}{|\mathcal{C}|}\sum_{c\in\mathcal{C}}F1_c.
\end{align}
Balanced accuracy averages per-class recall and reflects how well the model recognizes each class under class imbalance:
\begin{align}
\mathrm{balanced\_acc} &= \frac{1}{|\mathcal{C}|}\sum_{c\in\mathcal{C}}R_c.
\end{align}

\paragraph{FCR metrics.}
For fault chain reasoning, we evaluate structural correctness from three perspectives: node recovery, edge recovery, and exact chain match.

For sample $i$, define the node sets
\begin{align}
G_i=\mathrm{set}(g_i), \qquad P_i=\mathrm{set}(p_i),
\end{align}
and the edge sets induced by adjacent node pairs:
\begin{align}
E(g_i)=\{(g_{i,j},g_{i,j+1})\}_{j=1}^{|g_i|-1},\quad
E(p_i)=\{(p_{i,j},p_{i,j+1})\}_{j=1}^{|p_i|-1}.
\end{align}

Node precision (node\_p) measures how many predicted nodes are correct:
\begin{align}
\mathrm{node\_precision}_i &=
\begin{cases}
\frac{|G_i\cap P_i|}{|P_i|}, & |P_i|>0,\\
0, & |P_i|=0,
\end{cases}
\end{align}
Node recall (node\_r) measures how much of the ground-truth chain is covered:
\begin{align}
\mathrm{node\_recall}_i = \frac{|G_i\cap P_i|}{|G_i|}.
\end{align}
Edge precision (edge\_p) measures the correctness of predicted causal transitions:
\begin{align}
\mathrm{edge\_precision}_i &=
\begin{cases}
\frac{|E(g_i)\cap E(p_i)|}{|E(p_i)|}, & |E(p_i)|>0,\\
0, & |E(p_i)|=0,
\end{cases}
\end{align}
Exact chain match (chain\_em) evaluates whether the whole predicted chain exactly matches the ground truth:
\begin{align}
\mathrm{chain\_em}_i = \mathbf{1}[g_i=p_i].
\end{align}

Dataset-level FCR metrics are macro averages over all samples.

\paragraph{FCR module-level ranking metrics.}
\label{FCR_module_metrics}
The following metrics are used in \S\ref{exp:FCR module} to evaluate how effectively the multi-stage FCR pipeline concentrates the ground-truth (GT) chain toward the top of the candidate ranking.

Let $K$ denote the total number of candidate chains retrieved for a given sample, and let $r_i\in\{1,\ldots,K\}$ denote the rank position of the GT chain for sample $i$ ($r_i=\infty$ if the GT chain is absent from the candidate set). Then:

\textit{Mean Reciprocal Rank} (MRR) measures how close the GT chain is to the top of the ranking on average:
\begin{align}
\mathrm{MRR} = \frac{1}{M}\sum_{i=1}^{M}\frac{1}{r_i}.
\end{align}

\textit{Top-$K$ AUC} accumulates rank-position indicators across the top-$K$ slots to capture the overall concentration of the GT chain within the highest-ranked candidates:
\begin{align}
\mathrm{AUC@K} = \frac{1}{M}\sum_{i=1}^{M}\sum_{k=1}^{K}\mathbf{1}[r_i\leq k].
\end{align}

\textit{Recall at rank $k$} (R@$k$) measures the fraction of samples for which the GT chain appears within the top-$k$ positions:
\begin{align}
\mathrm{R@}k = \frac{1}{M}\sum_{i=1}^{M}\mathbf{1}[r_i\leq k].
\end{align}

\textit{Top-$K$ hit rate} is the cumulative recall curve reported in Fig.~\ref{fig:FCR_cumulative_hit_rate}. For each cutoff $k\in\{1,\ldots,K\}$, it equals R@$k$ and measures the fraction of test samples whose GT chain is ranked within the top $k$ after the final adjustment stage.


\section{Baselines}\label{appen:baselines}
We compare \textit{QoEReasoner} against baselines spanning four paradigms: rule-based heuristics, deep learning, pure LLMs, and LLM-based agents.

\paragraph{Heuristic Baseline.}
The rule-based baseline is an expert-engineered deterministic diagnostic workflow. For each KPI time-series window, a set of hand-crafted conditional rules checks whether specific KPI combinations exceed predefined thresholds.

\paragraph{Deep Learning Baselines.}
\begin{itemize}[itemsep=0em, topsep=0em, leftmargin=1em]
    \item \textbf{TS-TCC}~\cite{eldele2021ts2vec}: a contrastive representation learning framework for time-series data. It learns informative representations from augmented time-series inputs, which are then used for downstream classification.
    
    \item \textbf{CA-TCC}~\cite{eldele2023self}: a semi-supervised extension of TS-TCC for few-labeled settings. It combines representation learning on unlabeled time-series data with class-aware optimization, and uses the learned representations for downstream classification.
\end{itemize}

\paragraph{Pure LLM Baselines.}
For each LLM, the raw or summarized KPI time-series data, together with a task-specific prompt, is directly provided to the model for diagnosis. No external tools, domain knowledge bases, or iterative reasoning loops are used.
\begin{itemize}[itemsep=0em, topsep=0em, leftmargin=1em]
    \item \textbf{DeepSeek-V3.2}~\cite{liu2025deepseek}: a recent large-scale Mixture-of-Experts language model with strong general reasoning ability, used here as a representative open-weight LLM backbone.
    
    \item \textbf{GPT-5.2}~\cite{hurst2024gpt}: a frontier proprietary LLM from OpenAI with strong instruction-following and reasoning capabilities, representing a state-of-the-art commercial LLM API.
    
    \item \textbf{Qwen3-14B}~\cite{yang2025qwen3}: a medium-scale open-weight LLM from the Qwen3 family, included to evaluate whether smaller models can remain competitive in domain-specific diagnosis tasks.
\end{itemize}

\paragraph{Agent Baseline.}
\textbf{RCA-Agent}~\cite{xu2025openrca} is a ReAct-style root cause analysis agent. It supports multi-step reasoning and iterative evidence gathering over heterogeneous observability signals through a controller--executor workflow built on top of an LLM backbone.

\section{Rubrics of User Study}
\label{appen:user-study}

The expert evaluators assessed each report along four dimensions: Correctness, Evidence Grounding, Knowledge Grounding, and Interpretability. Each dimension was rated on a 5-point Likert scale, with higher scores indicating better quality.

\begin{itemize}[itemsep=0em, topsep=0em, leftmargin=1em]
    \item \textbf{Correctness}
    \begin{itemize}[itemsep=0em, topsep=0em, leftmargin=1em]
        \item \textbf{1:} Neither the main causal chain nor the alternative candidate chains are valid, and the causal direction is clearly incorrect.
        \item \textbf{2:} Mentions a few relevant variables, but fails to form an effective causal chain.
        \item \textbf{3:} Contains some correct chain segments, but the overall chain is incomplete, or key intermediate nodes are incorrect.
        \item \textbf{4:} The main chain is mostly correct, and the major intermediate steps are reasonable, but there are minor jumps or weak links.
        \item \textbf{5:} The main chain is complete and correct, with a clear and valid progression from root cause to intermediate mechanism to observed phenomenon/result.
    \end{itemize}

    \item \textbf{Evidence Grounding}
    \begin{itemize}[itemsep=0em, topsep=0em, leftmargin=1em]
        \item \textbf{1:} Almost no data support is provided; the report mainly states conclusions without evidence.
        \item \textbf{2:} Refers to a small amount of KPI data, but the connection to the conclusions is weak, or only supports a small part of the chain.
        \item \textbf{3:} Key conclusions are partially supported by data, but there are unverified jumps in the intermediate reasoning chain.
        \item \textbf{4:} Most key nodes are supported by data, and both supporting and opposing evidence are considered.
        \item \textbf{5:} The report provides explicit data support from root cause to intermediate steps to final outcomes, and clearly explains the strength of evidence as well as remaining gaps.
    \end{itemize}

    \item \textbf{Knowledge Grounding}
    \begin{itemize}[itemsep=0em, topsep=0em, leftmargin=1em]
        \item \textbf{1:} The reasoning clearly contradicts domain knowledge, or the meanings of KPIs are used incorrectly.
        \item \textbf{2:} Contains multiple unreasonable mappings between KPIs and mechanisms, and the explanations are forced or implausible.
        \item \textbf{3:} The overall perspective is reasonable, but some mechanistic chains are oversimplified or lack rigor.
        \item \textbf{4:} Consistent with most domain knowledge, and the mapping between KPIs and fault mechanisms is largely correct.
        \item \textbf{5:} The mechanistic explanation is accurate and rigorous, and it clearly distinguishes direct evidence, indirect evidence, and counter-evidence.
    \end{itemize}

    \item \textbf{Interpretability}
    \begin{itemize}[itemsep=0em, topsep=0em, leftmargin=1em]
        \item \textbf{1:} Confusing and difficult to understand. The output lacks basic organization, with information piled together chaotically; phenomena, conclusions, evidence, and suggestions are hard to distinguish; reviewers cannot quickly understand what the agent is trying to convey.
        \item \textbf{2:} Barely readable, but loosely structured. Some structure is present, but the organization is unstable; conclusions, reasoning, evidence, and suggestions are mixed together; the content can be roughly understood, but reviewers must reorganize the information themselves.
        \item \textbf{3:} Basically clear in structure. The report has a reasonably clear organization and can distinguish major conclusions, some evidence, and some reasoning; it is generally understandable, but may still contain repetition, unnatural ordering, or insufficient emphasis on key points.
        \item \textbf{4:} Clear hierarchy with distinct priorities. The main conclusion is prominent; causal chains, evidence, counter-evidence/limitations, and suggested actions are clearly separated; the writing is fairly concise, with no obvious redundancy, allowing reviewers to read and understand it efficiently.
        \item \textbf{5:} Highly clear and easy to review and verify. The output is very well structured, with the main conclusion immediately visible; causal chains, evidence, counter-evidence, uncertainties, and recommended actions are clearly aligned; the language is concise and information-dense, enabling different reviewers to quickly understand, compare, and validate the report.
    \end{itemize}
\end{itemize}




\end{document}